# DSRC-Enabled Train Safety Communication System at Unmanned Crossings

Junsung Choi, *Member*, *IEEE*, Vuk Marojevic, *Senior Member*, *IEEE*, Carl B. Dietrich, *Senior Member*, *IEEE*, and Seungyoung Ahn, *Senior Member*, *IEEE*

***Abstract*— Although wireless technology is available for safety-critical applications, few applications have been used to improve train crossing safety. To prevent potential collisions between trains and vehicles, we present a Dedicated Short-Range Communication (DSRC)-enabled train safety communication system targeting to implement at unmanned crossings. Since our application's purpose is preventing collisions between trains and vehicles, we present a method to calculate the minimum required warning time for head-to-head collision at the train crossing. Furthermore, we define the best- and worst-case scenarios and provide practical measurements at six operating crossings in the U.S. with numerous system configurations such as modulation scheme, transmission power, antenna type, train speed, and vehicle braking distances. From our measurements, we find that the warning application coverage range is independent of the train speed, that the omnidirectional antenna with high transmission power is the best configuration for our system, and that the latency values are mostly less than 5 ms. We use the radio communication coverage to evaluate the time to avoid collision and introduce the safeness level metric. From the measured data, we observe that the DSRC-enabled train safety communication system is feasible for up to 35 mph train speeds which is providing more than 25-30 s time to avoid the collision for 25-65 mph vehicle speeds. Higher train speeds are expected to be safe, but more measurements beyond the 200 m mark with respect to a crossing considered here are needed for a definite conclusion.**

## I. Introduction

The ongoing effort to increase road safety and reduce accidents will benefit from Intelligent Transportation Systems (ITSs). Road safety must include safer railroad crossings. A driver's lack of awareness of an approaching train can cause collisions between trains and vehicles which can lead to injury or death. About 85% of all reported railroad accidents between 2011 to 2020 were due to train-to-vehicle collisions and most of these accidents happened near crossings [1]. Even though the efforts of projects such as Operation Lifesavers, which focus on installing improved safety features at railroad crossings, 80% of U.S. crossings are still classified as *unprotected*, meaning that there are no lights, warnings, or gates [1].

While mechanical barriers, such as gates would certainly improve safety, advances in wireless communications can be leveraged to provide a scalable solution. Dedicated Short-Range Communications (DSRC) [2] has been the main protocol for vehicle-to-vehicle (V2V) and vehicle-to-infrastructure (V2I) communications, and a key technology enabler for Intelligent Transportation Systems (ITSs). DSRC operates at 5.9 GHz with seven 10 MHz channels between 5.850 and 5.925 GHz as shown in Fig. 1. One DSRC safety application is early warning, which is meant to avoid vehicle collisions [3-9].

Unlike a typical V2V or V2I communication environment, a train-related communication environment has different constraints. The communication environment near a crossing is similar to intersections on a vehicular road; however, the visibility of trains and vehicles is much narrower. Narrower visibility relates to reduced preparation time to stop which can cause accidents. In addition, the time to make a complete stop is different because trains need much longer than vehicles to come to a full stop; in other words, vehicles are able to react more effectively to avoid collisions.

In this paper, we present an architecture for communications at unprotected railroad crossings to enable early warning to vehicles of an approaching train. Because of the difficulty of a train driver to react and stop the train, the train's role is that of a transmitter and the vehicle's role is that of a receiver in the architecture. The proposed communication system is also considering the situation of a vehicle not being equipped with a DSRC radio. The proposed architecture is using an infrastructure node near the crossing to act as the DSRC receiver that issues a warning or relays the message. The proposed architecture, therefore, combines V2V and V2I communication. Because trains are running on known tracks,

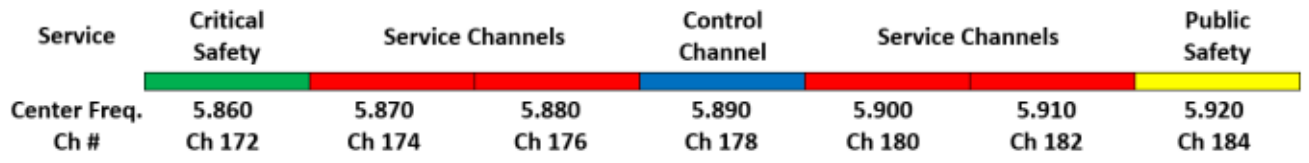

Fig. 1. DSRC channel allocations in the U.S.

J. Choi and S. Ahn are with the CCS Graduate school of Green Transportation, Korea Advanced Institute of Science and Technology, Daejeon 34051, Korea (e-mail: choijs89; sahn@kaist.ac.kr)

V. Maroejevic is with the Department of Electrical and Computer Engineering, Mississippi State University, Starkville, MS 39752, USA (e-mail: vuk.marojevic@msstate.edu)

C. B. Dietrich is with the Bradley Department of Electrical and Computer Engineering, Virginia Tech, Blacksburg, VA 24060, USA (e-mail: cdietric@vt.edu)

the DSRC receiver near the crossing determines whether and when to trigger a warning based on the reported location of the train to prevent a potential collision.

We conduct DSRC performance measurements under a variety of train, vehicle, and system operational conditions. Our measurements are processed at crossings of the Transportation Technology Center, Inc. (TTCI) at Pueblo, Colorado, and the Shenandoah Valley Railroad (SVRR), a Class 3 short-line railroad in Staunton, Virginia. The open space and artificial shadowing environments, which we consider as the best- and worst-case scenarios that the communication system may face, are conducted at the TTCI. Measurements representing suburban and rural environments are conducted at the SVRR.

In this paper, we present the feasibility of adapting DSRC for the proposed train safety communication system. We suggest safety requirements in the form of head-to-head collision by defining the safeness level metric, which depends on the position of the train, the speeds of the train and vehicles, and the radio coverage. From the evaluations of DSRC measurements in the best- and worst-case settings and six different operational crossings, we obtain the DSRC coverage range and system latency, which are directly converted to the necessary components to define the safeness level. In this paper, we do not provide statistical data, but present a practical feasibility analysis based on measurement in a variety of production scenarios. We have not studied the capacity problem of V2X in this work, which is an issue in dense vehicular environments. For the proposed application, it is likely not an issue, but the use of DSRC for V2V and train-to-vehicle (T2V) combined is appealing and needs to be analyzed.

The rest of the paper is organized as follows: Section II discusses the related research and highlights our contributions. Section III describes the safety communication system architecture, the safety requirements, and the safeness level metric. Section IV explains the measurement methodology. Sections V and VI present the TTCI and SVRR results and evaluations. Section VII draws the conclusions.

## II. Related Work and Contributions

Warning applications have been studied in several contexts, including V2V and V2P, assuming Wi-Fi and Long-Term Evolution (LTE)-based system. In [7], the authors combine IEEE 802.11p-based multi-hop clustering and a fourth-generation (4G) cellular system. In [9], the authors compare the performance between IEEE 802.11p and LTE for vehicles (LTE-V) for the intersection collision warning system (ICWS), which is an environment similar to our warning application. In addition, it has a similar architecture and uses a roadside unit (RSU) placed near the intersection acting as a relay. The concept is also shown in [10], which presents a V2V relay network in an obstructed environment. References [7]–[11] discuss methodologies, major components, architectures, and minimum requirements for collision warning services.

Studies have been done for using advanced wireless technologies, such as satellite, WiFi, optical wireless communications (OWC), LTE, and GSM-R for train communications services [12, 13]. The authors of [13] provide a technical survey of possible candidates for railroad technologies, including GSM-R, Wi-Fi, WIMAX, LTE-R, RoF, LCX, and cognitive radio. They conclude that WIMAX and LTE could be the best technologies for train-related communications; interestingly, the authors did not consider DSRC as a candidate. Reference [14] presents an LTE-railroad (LTE-R) testbed with an Internet protocol (IP)-based network architecture. The authors demonstrate the performance of the testbed and confirm that reliable communications and multimedia services that require high data rates are feasible. The design of a train-to-ground communication system using LTE technology (LTE-M) for urban rail transit/metro with advantages and disadvantages of LTE-M through field trials are summarized in [15]. Similarly, train-ground communication systems for urban rail transit using Time Division-LTE (TD-LTE) is studied in [16]. In [17], the authors introduce a GSM-R-based train-related monitoring system for the entire rail network in the UK. The authors of [18] introduce free-space-optics (FSO) coverage models for high-speed-train communications. As shown in multiple studies, there are numerous wireless technologies that can be adapted for train-related communications. Also, most of the prior works focus on a communication architecture that is either train-to-ground [15, 16, 18, 19] or train-to-train [17, 20, 21]. In this paper, we analyze both train-to-ground and train-to-vehicle communications using the DSRC protocol.

The concept of T2V communications was introduced in [22], which describes a train-to-vehicle early warning system designed and managed for trains and vehicles near railroad crossings in Australia. There, the authors are more focused on the warning architecture and on improving the degree of noticing warnings than on the performance of the radio communications system; without considering the message transmission and reception performance, the performance of the proposed method will depend on the conditions where the application is implemented.

Practical studies related to T2V communications exist. In [23] and [24], the authors explore the antenna criteria for a railroad crossing safety application and derive an optimal antenna pattern for practical installations. The authors of [25-27] introduce a T2V early warning system architecture and present their DSRC performance and propagation channel measurements at the measurement sites. This paper extends our early works [25-27]. In those works, the DSRC technology has been shown to have the potential to offer a cost-effective approach to the deployment of an early warning system to avoid train-to-vehicle collisions.

In this paper, we present a more detailed DSRC-enabled early warning system architecture introducing new train-to-ground and train-to-vehicle communications constraints and analyses. We elaborate the head-to-head collision safety requirements at a crossing while many prior works are considering head-to-tail collision safety [4, 5, 20, 21]. Also, we present new DSRC performance results and evaluation of six additional crossings. Since our application of interest is that of a radio-enabled warning system, we consider not only the radio performance, but also the practical implications and, specifically, consider the absolute number of correctly received packets. The analysis of the communication and timing performance is done with respect to the safety requirements that we introduce. We do not provide statistical data, but present a practical feasibility analysis through measurement obtained in a variety of production scenarios.

## III. Train Safety Communication Architecture and Safety Requirements

### A. *Train Safety Communication Architectures*

The proposed train safety communication system at unprotected crossings has the following components: an approaching train, an approaching vehicle, and fixed infrastructure, such as a signal pole near the crossing. The train safety communication system operates in two cases, *direct warning* and *indirect warning*. Both are illustrated in Fig. 2.

For the *direct warning* case, shown in Fig. 2(a), the vehicle driver receives the warning directly from the train. This scenario is the preferred one to implement where the radio channel between the train and vehicles has a strong line-of-sight (LoS) component. For the *indirect warning* case, illustrated in Fig. 2(b), the infrastructure at the railroad crossing receives the warning from the train and possibly retransmits it to the vehicles or generates a warning in the form of light or sound for the vehicles that do not have DSRC radios. The *indirect warning* case is preferred where the propagation channel between the infrastructure and the train has a stronger LoS component than that between the train and vehicles near the crossing. The *direct warning* case corresponds to a T2V communication system and the *indirect warning* case corresponds to a train-to-infrastructure (T2I) system.

### B. *Safety Requirements*

To avoid a collision, the vehicle driver should notice the coming train before either the vehicle or train reaches the crossing. Since it is much harder for the train to make a complete stop than for a vehicle, we consider the vehicle driver being notified of an approaching train and taking the action to avoid a collision. In order to evaluate the safety protocol, we must know the time that remains for the train to reach the crossing ($t_t$), which can be written as:

$$t_t = \frac{d_t}{v_t}, \tag{1}$$

where $v_t$ represents the train speed and $d_t$ the remaining distance between the train and the crossing. Parameter $t_t$ depends on the instantaneous train position and speed.

The time to avoid collision ($t_{TAC}$) is the time that the vehicle driver notices the warning and reacts to it ($t_r$) plus the system delay ($t_s$), the braking time ($t_b$), and a certain protection time ($t_{prot}$), which depends on the radio communication performance. Fig. 3 illustrates these parameters. We study the warning system for vehicles that have not already entered the crossing. A DSRC receiver comes with a Global Positioning System (GPS) receiver. Thus, $t_{TAC}$ is a function of the train speed and the remaining train distance to the crossing ($d_{warn}$):

$$t_{TAC} = \frac{d_{warn}}{v_t} = t_r + t_s + t_b + t_{prot}. \tag{2}$$

More precisely, $d_{warn}$ refers to the maximum distance between the train and the crossing where the DSRC-based warning is successful and reliable, i.e. where the number of received packets by the DSRC receivers is above a certain threshold. It depends on the transmission power level, modulation scheme, antennas, and RF environment and can be empirically obtained as we will show in this paper. We measure it for a single receiver located near the crossing (indirect warning case) or at a certain distance on the road toward the crossing (direct warning case). Note that $d_{warn}$ here is crossing specific and that a single crossing may have different $d_{warn}$ instances for different radio configurations.

With larger $t_{prot}$, the driver has more time to avoid a potential collision. The nominal $t_r$ is 3.5 s [21]. The system delay, $t_s$, has two components: the wireless propagation delay, $t_{prop}$, and the system processing time, $t_{proc}$:

$$t_s = t_{prop} + t_{proc}. \tag{3}$$

For the *direct warning* case, $t_{prop}$ is the propagation delay between the train and the vehicle, whereas $t_{proc}$ is the processing time at the vehicle receiver. For the *indirect warning* case, $t_{prop}$ needs to be considered between the train and the infrastructure and between the infrastructure and the vehicle.

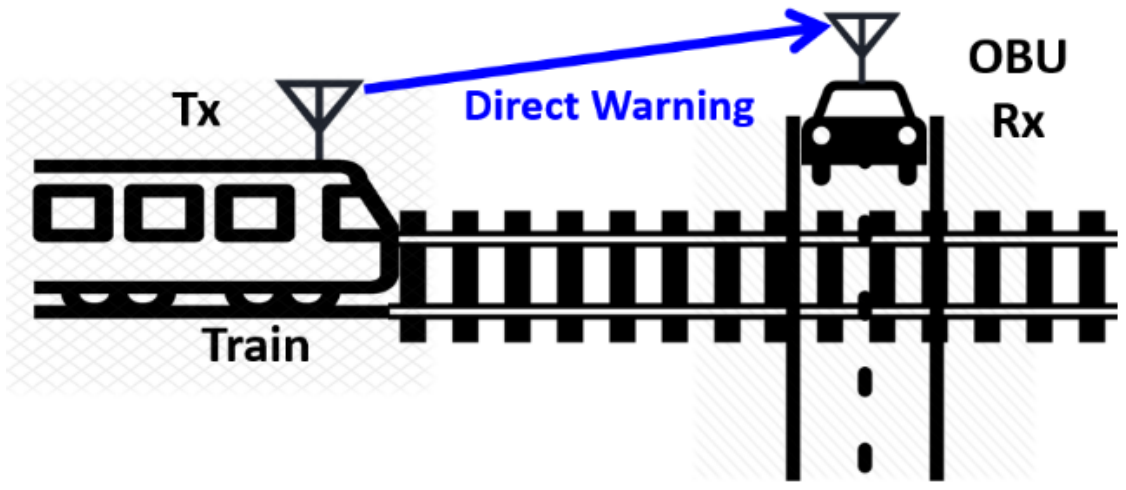

(a)

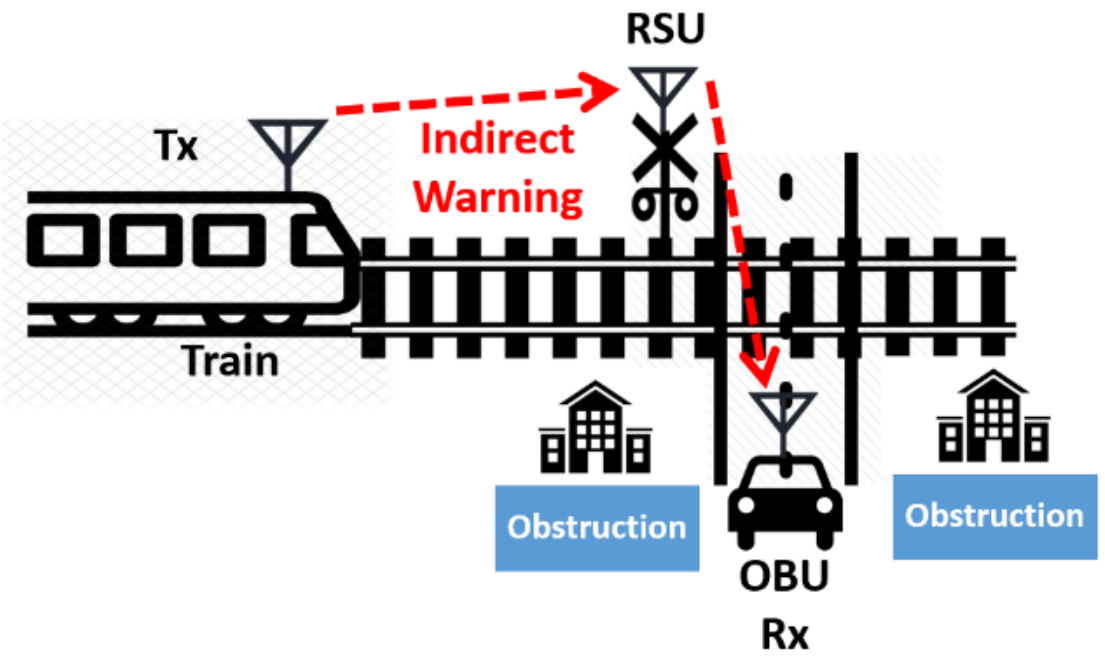

(b)

Fig. 2. Train safety communication system cases: (a) direct warning; (b) indirect warning.

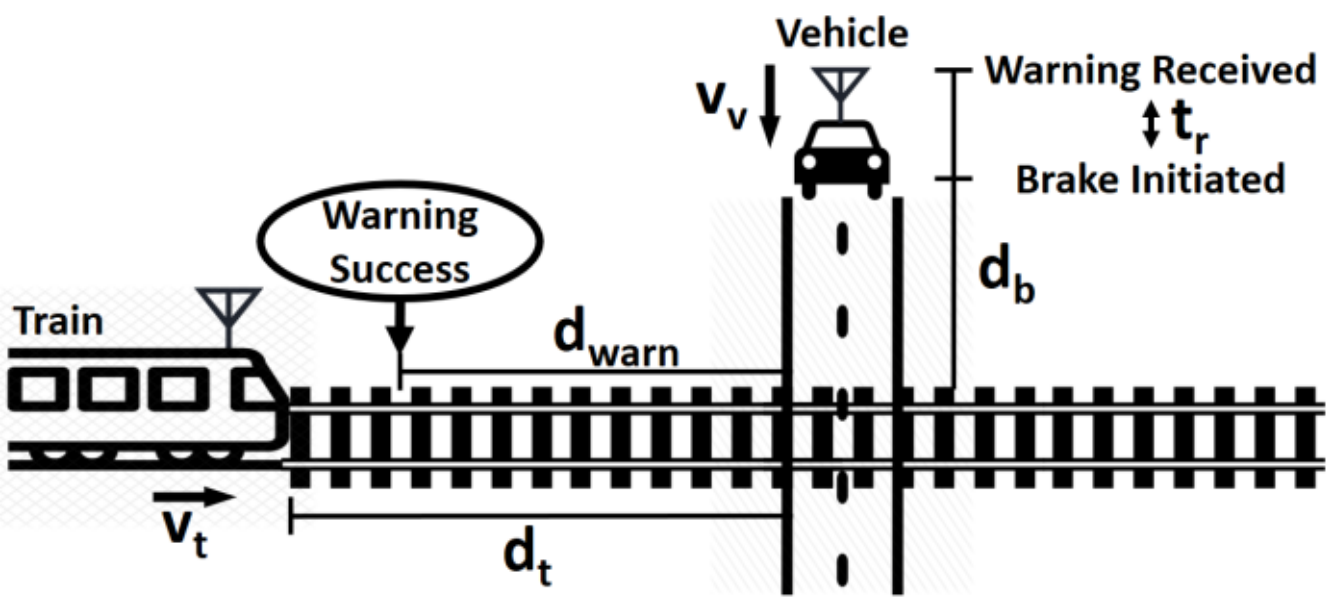

Fig. 3. Safety requirement parameters.

TABLE I
AVERAGE VEHICLE STOPPING DISTANCE AS A FUNCTION OF SPEED [28]

| Vehicle speed (mph) | Vehicle speed (m/s) | $d_b$ on dry road (m) | $t_b$ on dry road (s) | $d_b$ on wet road (m) | $t_b$ on wet road (s) |
|---|---|---|---|---|---|
| 25 | 11.11 | 25.5 | 2.3 | 51.3 | 4.62 |
| 35 | 15.55 | 41.4 | 2.66 | 82.8 | 5.32 |
| 45 | 20.00 | 59.1 | 2.96 | 118.2 | 5.91 |
| 55 | 24.44 | 79.8 | 3.27 | 159.3 | 6.52 |
| 65 | 28.89 | 103.2 | 3.57 | 206.7 | 7.15 |

Correspondingly, $t_{proc}$ needs to also account for the processing at the RSU that relays the signal to the vehicles.

Parameter $t_{prop}$ varies with $d_t$ and the remaining distance between the vehicle and the crossing, $d_v$. Since the safety application uses electromagnetic signal transmission over distances in the order of one hundred meters, $t_{prop}$ will be less significant when compared to $t_{proc}$ and may be negligible. If the protection time, $t_{prot}$, results in a negative figure, the system fails and cannot provide collision safety.

The braking time, $t_b$, depends on the vehicle speed, $v_v$, and suggested braking distance, $d_b$, which is given in [28]. The usual vehicle speeds in the U.S. are 25, 35, 45, 55, and 65 mph and these are considered for calculating the corresponding braking distances in Table I. The braking distances, $d_b$, are in the range of 25.5-206.7 m leading to a $t_b$ of 2.3-7.15 s.

To evaluate the safeness in response to a moving train and road vehicle, the following components are required: $t_t$, $t_{TAC}$, $t_r$, $t_s$, and $t_b$. With these, the safeness level ($\psi$) is obtained as

$$\psi = \frac{t_t-(t_r+t_s+t_b)}{t_{TAC}-(t_r+t_s+t_b)} \, . \quad (4)$$

Parameter $\psi$ changes with the instantaneous train position, $d_t$, which is directly related to $t_t$, and the three different safety conditions can be defined as shown in Fig. 4. When $t_{TAC}$ is between the remaining time for the train to arrive at the crossing, $t_t$, and the vehicle driver reaction time, $t_r$, plus the remaining time for the vehicle to reach the crossing, $t_b$, we consider this as a '*no risk of collision*' scenario, characterized by $\psi$ being larger than or equal to 1. Figure 4(a) illustrates this. When $t_t$ is between $t_{TAC}$ and the sum of $t_r$ and $t_b$, as shown in Fig. 4(b), $\psi$ will be between 0 and 1, and we consider it still '*safe but close to colliding*'. When either $t_t$ or $t_{TAC}$ is smaller than the sum of $t_r$ and $t_b$ (Figs. 4(c) and (d)), $\psi$ is less than 0 and we consider it '*not safe*'; if $t_{TAC}$ is smaller than the sum of $t_r$ and $t_b$, the DSRC system fails to provide safeness, because the radio signals carrying the warning are decoded to late for vehicles who are about to cross the tracks to stop.

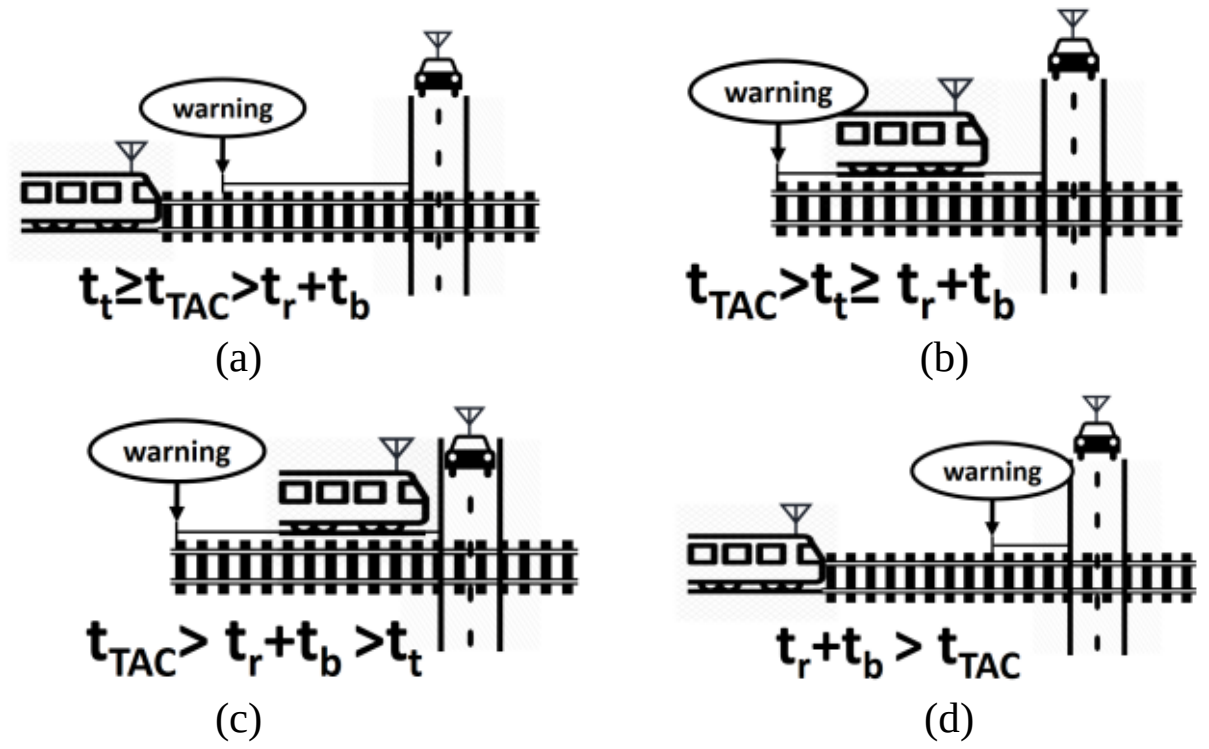

Fig. 4. Scenario examples for safeness level of (a) $\psi \geq 1$; (b) $1 > \psi \geq 0$; (c) $0 > \psi$; (d) $0 > \psi$ (system fails).

The safeness level categories are then as follows:

$$\begin{cases} \psi < 0; not\ safe \\ 0 \leq \psi < 1; safe\ but\ close\ to\ colliding. \\ \psi \geq 1; no\ risk\ of\ collision \end{cases} \quad (5)$$

From the time difference between where $\psi$ equals 0 and 1, we can evaluate the amount of the additional protection time for the vehicle driver to avoid the collision, $t_{prot}$, that the system can provide. The differences due to different train and vehicle speeds and notification time requirements require a proper warning message to be scheduled as a function of speed, geometry, and communications and other delays. The major parameters to consider for determining the safeness level (4) are the time of the train to reach the crossing (1), the time of the vehicle to reach the crossing which is the sum of $t_r$ and $t_b$, and at what point the warning succeded, $d_{warn}$, which can be converted to time, $t_{TAC}$, as shown in (2). With these main parameters which we can obtain from the measurable DSRC radio performance, we can evaluate the safeness level, $\psi$.

## IV. MEASUREMENT METHODOLOGY

### A. DSRC Performance Measurement Description

The purpose of the DSRC performance measurements and analysis is to evaluate the performance of the train safety communication system. The performance metric is the Packet Error Rate (PER) for a determined distance window. This distance is between the crossing and the instantaneous GPS location of the train. The PER is obtained by post-processing the data and comparing the received with the transmitted packets.

For proper system evaluation, we conduct the following measurement on a test track and in a production environment using the Cohda Wireless MK5 DSRC radios at the transmitter and receiver [29]. The best-case scenario that we consider is an unobstructed wide-open area, and the worst-case scenario is an obstructed (RF shadowing) environment. We conduct these scenarios with different speeds, 20-79 mph, at the TTCI test tracks. The production environment corresponds to 6 crossings along the SVRR tracks in Virginia. These 6 crossings show rural- or suburban-like environments with moving vehicles, buildings, and trees near the crossings. The measurements at the TTCI site provide the edge cases, whereas the SVRR site represents a typical production environment in the US.

### B. Measurement Parameters/Settings

#### 1) Testing Sites

The edge case measurements are conducted in two different sites along the TTCI tracks, which allows for high-speed train technology testing. As shown in Fig. 5(a), the best-case scenario is a wide-open space with no major obstructions near the test tracks. As shown in Fig. 5(b), the worst-case scenario is an artificial shadowing environment, where seven railroad cargos, of approximately 4 m height and 10 m length each, are parked on an adjacent track. The cargos are between the test track and the DSRC receiver so that the transmitter-receiver is mostly in non-line of sight (NLoS).

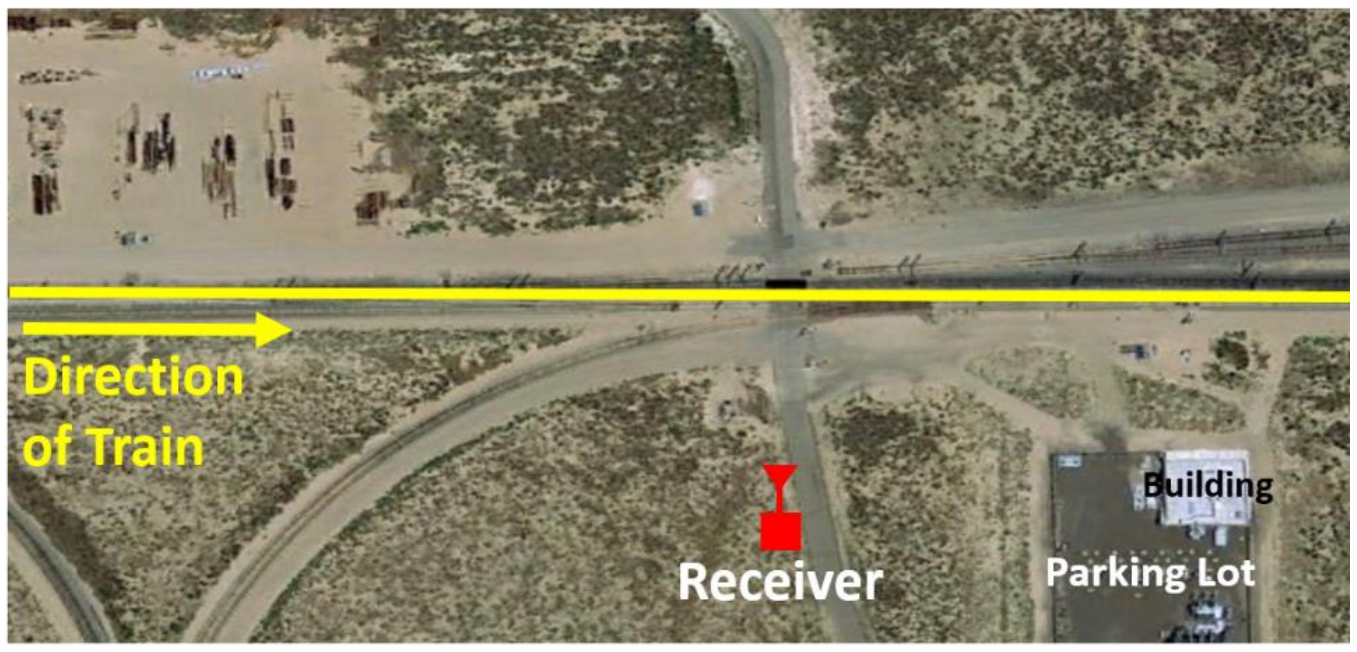

(a)

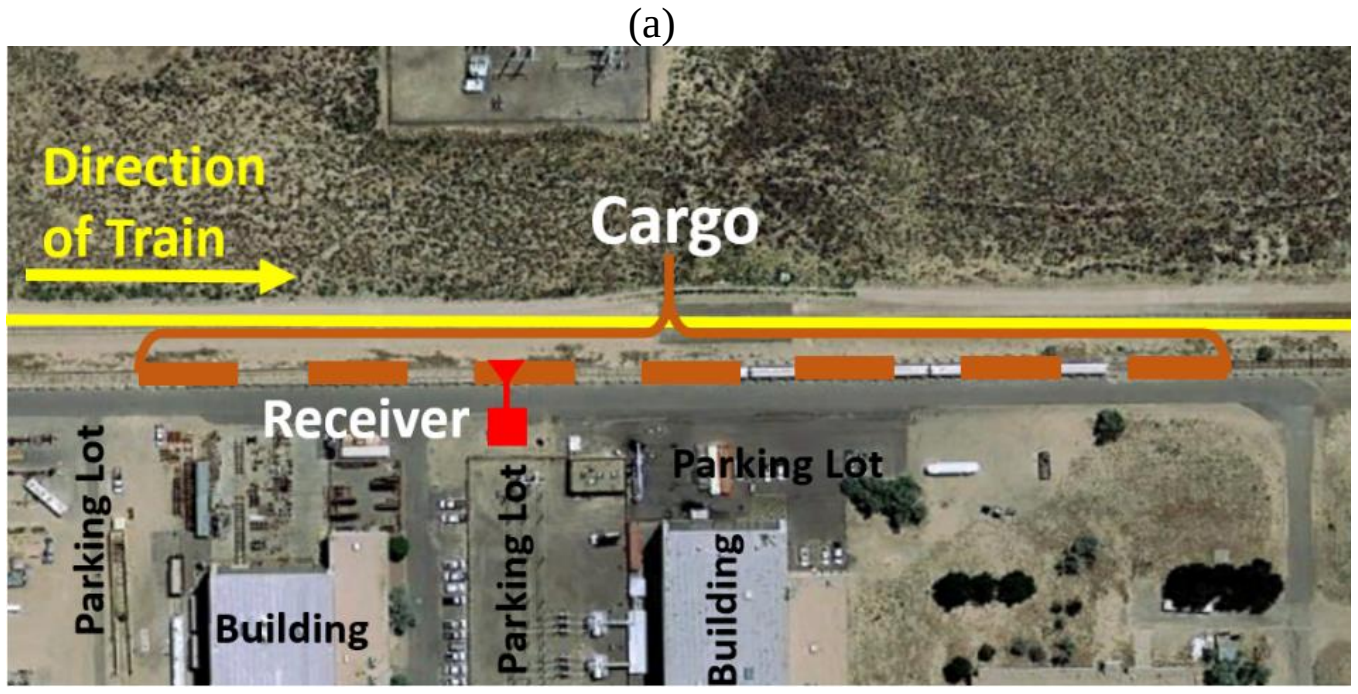

(b)

Fig. 5. TTCI track testing sites: (a) wide-open space; (b) artificial shadowing.

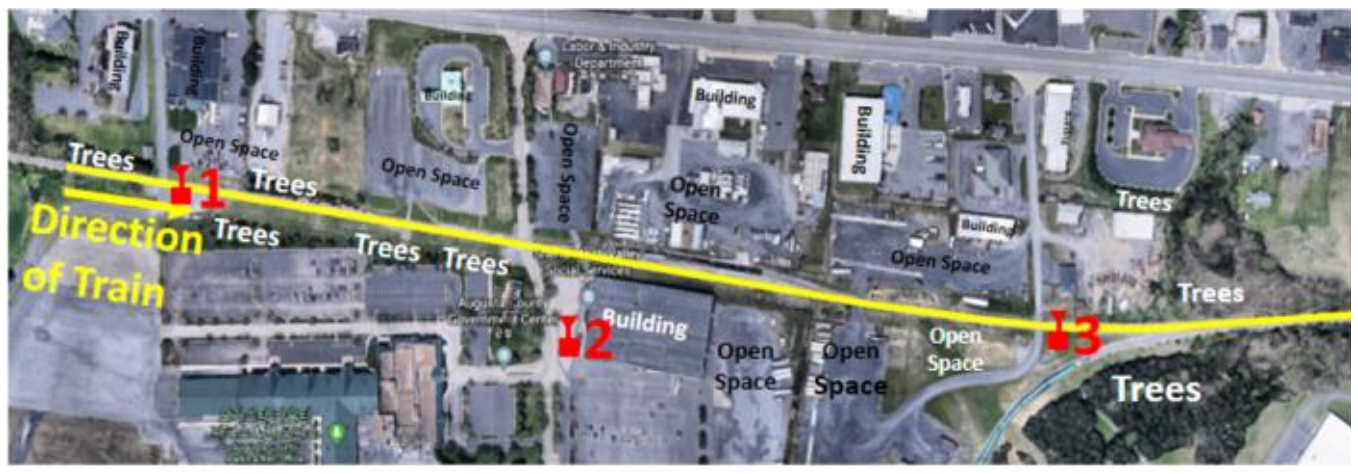

(a)

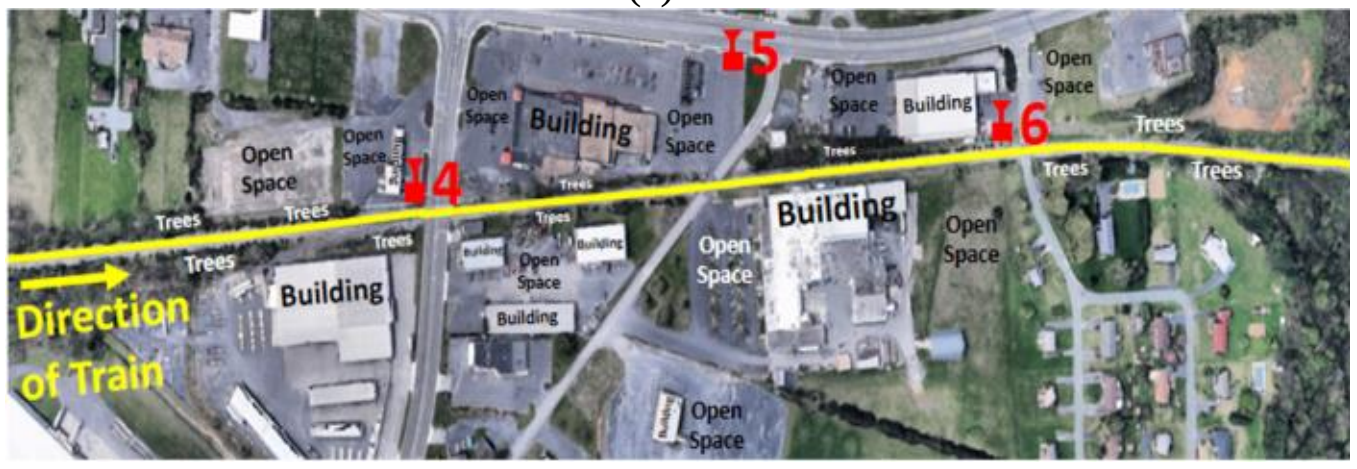

(b)

Fig. 6. SVRR track testing sites: (a) crossing #1-3; (b) crossing #4-6.

We consider six crossings along the SVRR tracks, as shown in Fig. 6. Fig. 6(a) depicts the crossings #1, #2, and #3 and Fig. 6(b) depicts the crossings #4, #5, and #6. We select crossings #1, #3, #4, and #6 for implementing the *indirect warning* case, thus DSRC RSUs are located at these crossings. We select crossings #2 and #5 for the *direct warning* case, and DSRC on-board units (OBUs) are placed on the road where an approaching vehicle should receive the warning. As shown in Fig. 6(a), the surroundings of crossings #1, #2, and #3 have a few buildings and open spaces resulting in mostly LoS between the train transmitter and the RSU/vehicle receivers. As shown in Fig. 6(b), the crossings #4, #5, and #6 are located in a suburban/urban type of environment where there are more buildings near the crossings. Crossing #4 is located right beside a four-lane vehicle road. At crossing #5, open space is behind a row of trees along the tracks with numerous buildings nearby.

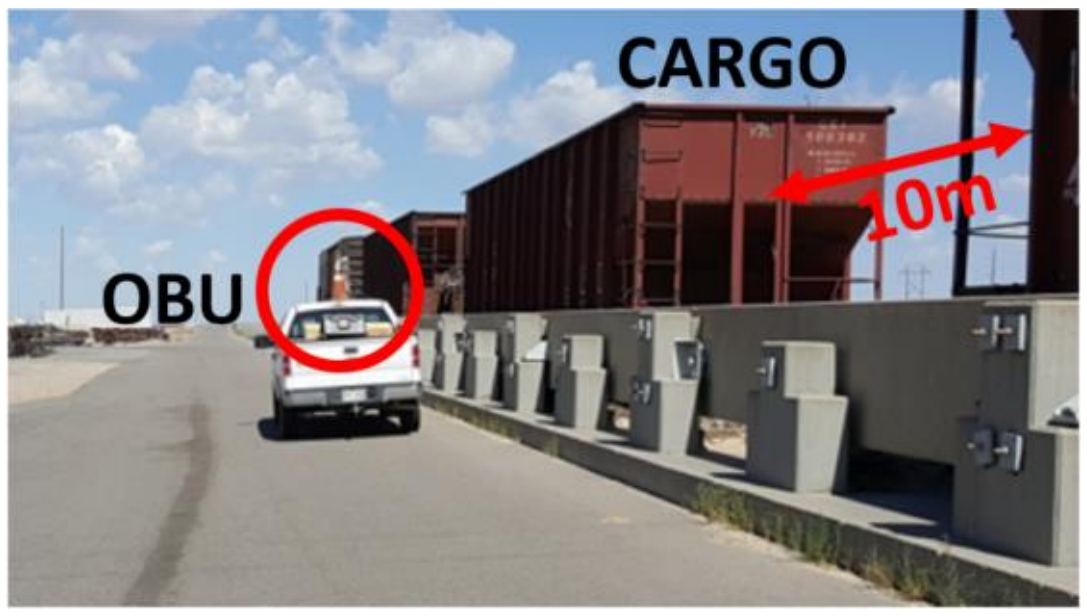

(a)

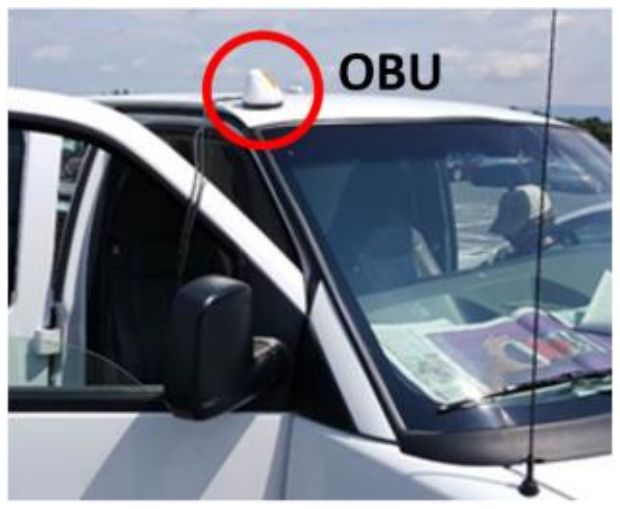

(b)

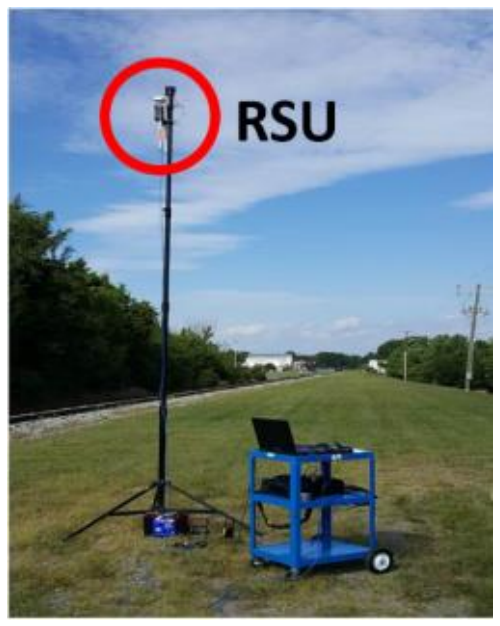

(c)

Fig. 7. DSRC receiver installation: (a) TTCI OBU setup; (b) SVRR OBU setup; (c) SVRR RSU setup.

### 2) *DSRC Receiver Positions*

DSRC OBUs are used for the TTCI measurements, whereas both DSRC OBUs and RSUs are used in the SVRR measurements. For both the TTCI and SVRR measurements, the antenna of the DSRC OBU receiver is on the roof of a vehicle at a height of about 1.7 m from the ground. This is shown in Fig. 7(a) for the TTCI track and in Fig. 7(b) for the SVRR track. For the wide-open space scenario on the TTCI track, the DSRC OBU is placed about 50 m from the crossing. For the artificial shadowing scenario, we placed the railroad cargos between the train's operating track and the DSRC OBU. The DSRC OBU is about 3 m from the railroad cargos, and the cargos are about 2 m from the operating track. For the SVRR track, we select crossings #2 and #5 as *direct warning* case with DSRC OBU receivers. The DSRC OBU at crossing #2 is about 42 m away from the track, and at crossing #5, it is about 36 m away from the track. The DSRC RSU receivers are hanging from the tip of a 3 m tall tripod, as shown in Fig. 7(c), and are placed about 5–7 m from the crossing.

### 3) *Configurations*

For both the TTCI and SVRR tracks, the DSRC OBUs and RSUs use a 6 dBi omnidirectional antenna [30]. The transmitter for the TTCI track uses a 12 dBi omnidirectional antenna [30] with a 23 dBm power level; both have similar antenna patterns [30]. The transmitter for the SVRR measurements uses a 12 dBi omnidirectional [30] and 23 dBi bidirectional antenna, which we built with two directional antennas with 10 degrees beamwidth on both the horizontal and vertical planes [31]. The transmit power level options are 11 and 23 dBm. The omnidirectional and bidirectional antenna patterns are shown in Fig. 8. The chosen modulation scheme is QPSK for the TTCI measurements, because of the more challenging setting, and QPSK or 16QAM for the SVRR measurements.

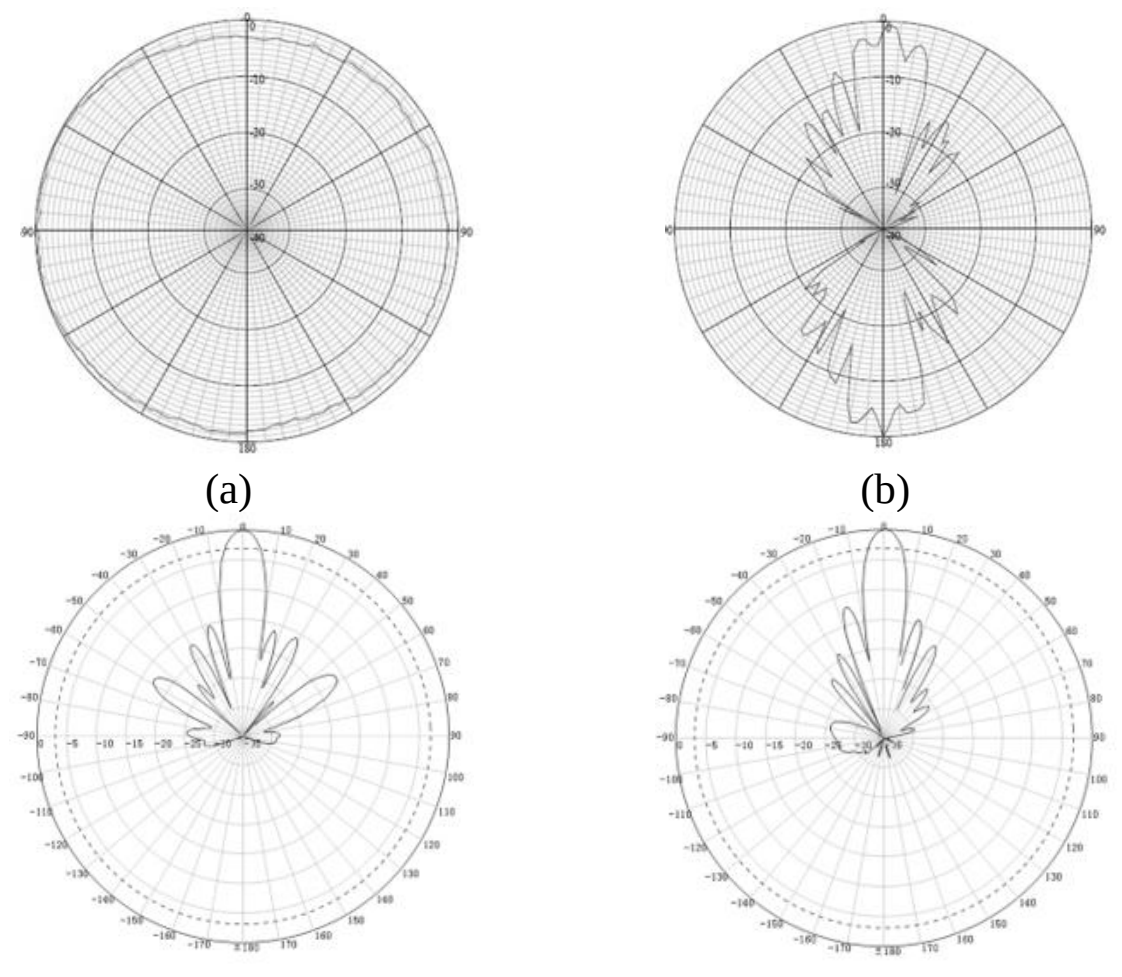

Fig. 8. Antenna pattern of (a) omnidirectional horizontal; (b) omnidirectional elevation; (c) bidirectional horizontal; (d) bidirectional elevation.

### 4) Train Operations

For both the TTCI and SVRR measurement, the train makes a pass through the testing region. For each pass, the train starts at about 1 km and 300–400 m distance from the crossing for the TTCI and SVRR measurements, respectively. The train accelerates to the desired test speed of 20, 50, and 79 mph for the TTCI measurements and 10 mph for the SVRR measurements. The speed of 79 mph is the highest speed for trains operating in the U.S., and 10 mph is the maximum speed permitted at the SVRR tracks. For all scenarios, the train maintains a constant speed through the test area before and after the crossing.

## C. Data Analysis Methods

We use the Basic Safety Message (BSM) application with a packet length of 99 bytes. The packets are continuously transmitted every 50 ms. The PER is calculated by comparing the transmitted and received packets with a 50 m range window for the TTCI measurement and a 20 m range window for the SVRR measurements. The average number of transmitted packets is 118, 47, and 29 in the 50 m window for the three speeds on the TTCI track and 94 in the 20 m window for the SVRR experiments. Since we consider the system for vehicular and passenger safety purposes, defining a proper threshold to evaluate the reliability is critical. We evaluate the reliability by comparing the number of correctly received packets with respect to the train position and assess the distance to the crossing where reliable packet reception occurs. Note that the BSM is transmitted as a short signal in the standard format where each transmission includes the updated GPS position of the train, the message count, the ID number, and the movement information such as acceleration, heading, and brake status.

The PER is analyzed as a function of the relative distance between the train and the crossing, where 0 m represents the crossing position and negative values represent the approaching train towards the crossing. When the number of correctly decoded packets at the DSRC receiver is higher than a certain threshold, we consider that the DSRC performance has *satisfied* the minimum requirement for the safety warning application. The threshold may differ for different applications. Because of the nature of the warning system, we consider the threshold being the successful reception of multiple packets before the crossing.

The system targets a moving train, which transmits BSMs at a given rate, which is 20 Hz, or one packet every 50 ms. The number of packets transmitted per distance traversed depends on the train speed. Even if the PER value is similar for different train speeds, the absolute numbers of correctly received packets will be different. This has implications on the reliability of generating warning messages. Each transmitted packet provides full information about a train's position, heading, and speed and the application can therefore inform the recipient of the message about the need for action with a single received packet. A reliable system should be able to receive multiple packets for redundancy and receive them early enough to ensure that the road vehicles are warned and react on time. Therefore, our definition of reliability is different from the typical DSRC performance evaluation metrics. We recommend considering not only the PER, but also the absolute number of correctly received packets for a given distance window. For this application, we are only interested in reliable communication before the train reaches the crossing. Table II summarizes the test configuration parameters.

TABLE II
DSRC RADIO AND TEST CONFIGURATIONS

| Configuration | *Testing Sites* | |
|---|---|---|
| | *TTCI* | *SVRR* |
| Center frequency | 5.87 GHz | 5.87 GHz |
| Channel Number | 174 | 174 |
| Modulation | QPSK | QPSK, 16QAM |
| Transmission Power | 23 dBm | 11, 23 dBm |
| Transmitter antenna | 12 dBi omnidirectional | 12 dBi omnidirectional, 23 dBi bidirectional |
| Receiver antenna | 6 dBi omnidirectional | 6 dBi omnidirectional |
| Train speed | 20, 50, 79 mph | 10 mph |
| Application | BSM | BSM |
| Packet size | 99 bytes | 99 bytes |

# V. TEST TRACK MEASUREMENT RESULTS AND RELIABILITY ANALYSIS

## A. DSRC Performance Results

Fig. 9 shows the DSRC performance results for the wide-open space and artificial shadowing scenarios. From Fig. 9(a), we find that there are almost no PER differences for all three speeds. The range for low PER after the crossing is shorter than before the crossing; we assume that the reason for this is the existence of an obstruction on the top of the locomotive located behind the antenna or the building next to the parking lot shown in Fig. 5(a), which may create NLoS conditions. From Fig. 9(b) we observe that the DSRC performance is clearly affected by the cargos causing shadowing. For all speeds, there is a significant performance degradation up to 150 m before the crossing. Experiments have shown that if LoS conditions can be provided during a packet transmission, the DSRC can successfully receive the transmitted packets. Between -150 m and 100 m, LoS links are possible due to the geometry and spacing between the cargo containers, leading to good DSRC performance. From both the wide-open space and artificial shadowing scenarios, we conclude that a LoS radio link is more relevant to the DSRC performance than the speed of the train. Higher speeds degrade the performance only slightly, as expected.

### B. *Reliability Analysis*

Fig. 10 indicates the number of correctly received packets for the wide-open space and artificial shadowing scenarios in 50 m intervals. The rate of transmission is constant which explains why more packets are received at lower train speeds.

As shown in Fig. 10(a), the number of received packets equals the number of transmitted packets for all three speeds. Also, more than 500 m coverage range is feasible for all three speeds.

Even though the PER values are lower for higher train speeds, as shown in Fig. 9(b), for the performance degraded region of -400 to -250 m, the number of correctly received packets are similar for all three speeds. This is captured in Fig. 10(b). We observe the increasing PER for slower speeds in this region. This explains that in terms of absolute numbers, the correctly received packets are similar. As a corollary, where the PER is zero, as in Fig. 9(a), more packets will be correctly received for the slower speed train, as observed in Fig. 10(a). Since several packets are correctly received independently of the train speed at -350 m, this is the figure for our reliability analysis.

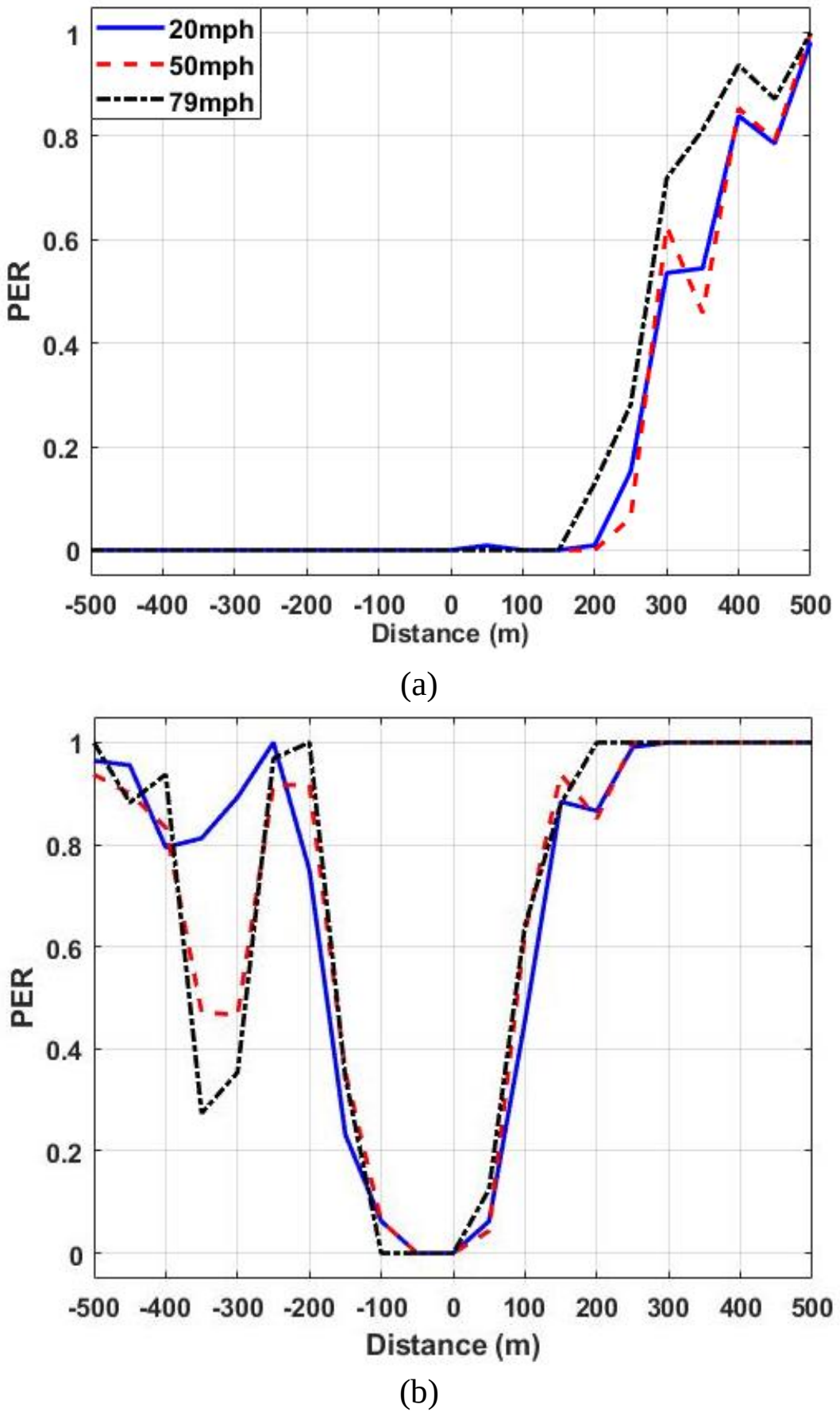

Fig. 9. DSRC PER results with respect to the relative train position with center at the crossing: (a) wide-open space; (b) artificial shadowing

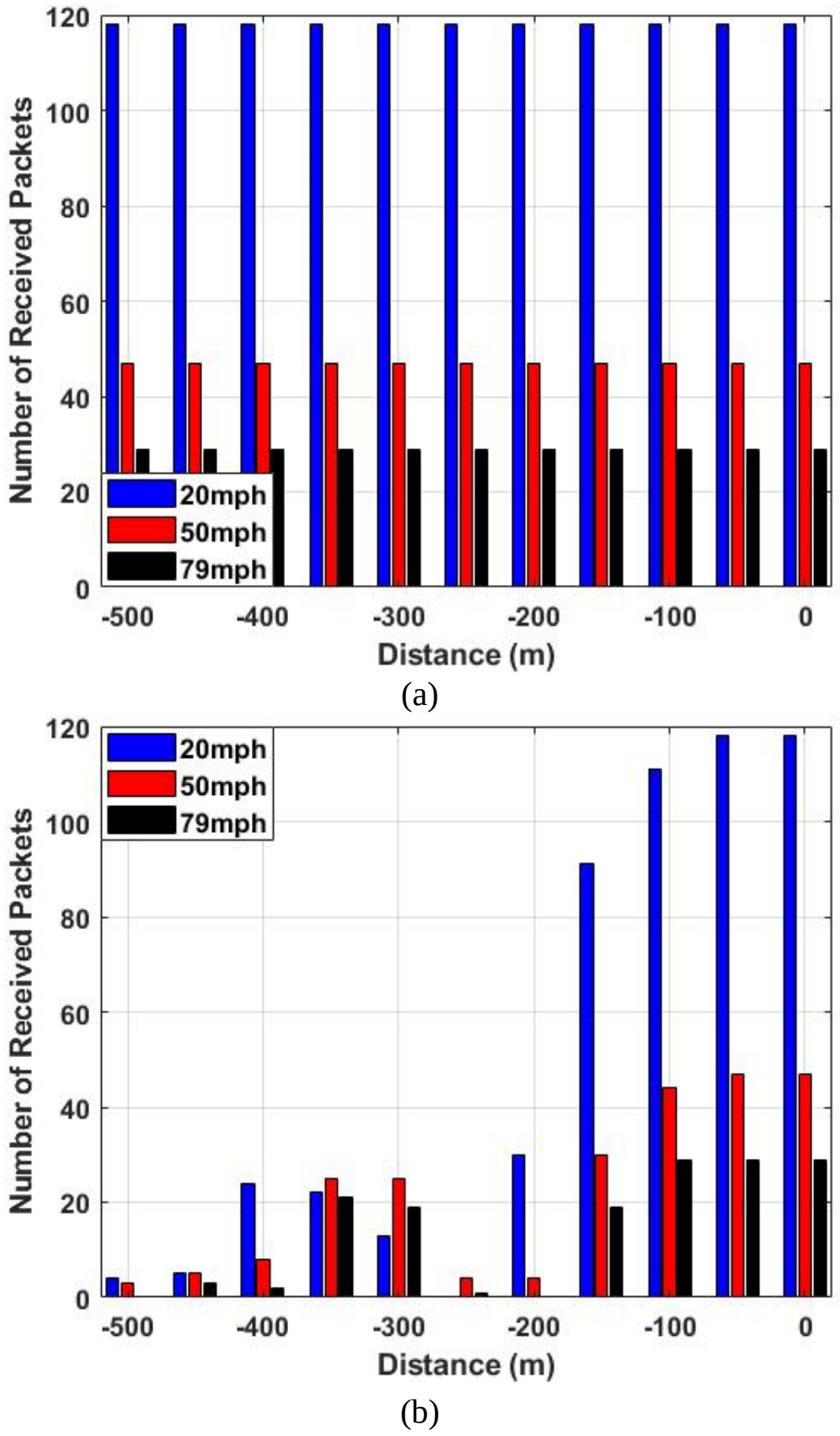

Fig. 10. Number of correctly received packets with respect to relative train position with center at the crossing: (a) wide-open space; (b) artificial shadowing

The successfully generated warning distance, $d_{warn}$, is 500 m before the crossing for the 20 and 50 mph train speeds and 450 m for the 79 mph train speed. The warning distance can be converted to times to avoid collision, $t_{TAC}$, which is calculated from (2). The times are 56, 22, and 12 s for the 20, 50, and 79 mph trains, respectively. Since these values are much larger than the sum of reaction time, $t_r$, and braking time, $t_b$, we conclude that the warning application can reliably operate for both edge use cases with the assumption that system delay, $t_s$, is small enough to be neglect.

## VI. Production Crossing Measurement Results and Reliability Analysis

Measurements are conducted at six crossings along the SVRR tracks in Staunton, Virginia. We designate crossings #1, #3, #4, and #6 for installing the proposed *indirect warning* case and crossings #2 and #5 for the *direct warning* case. The train operates at its maximum allowed speed from 200 m before the

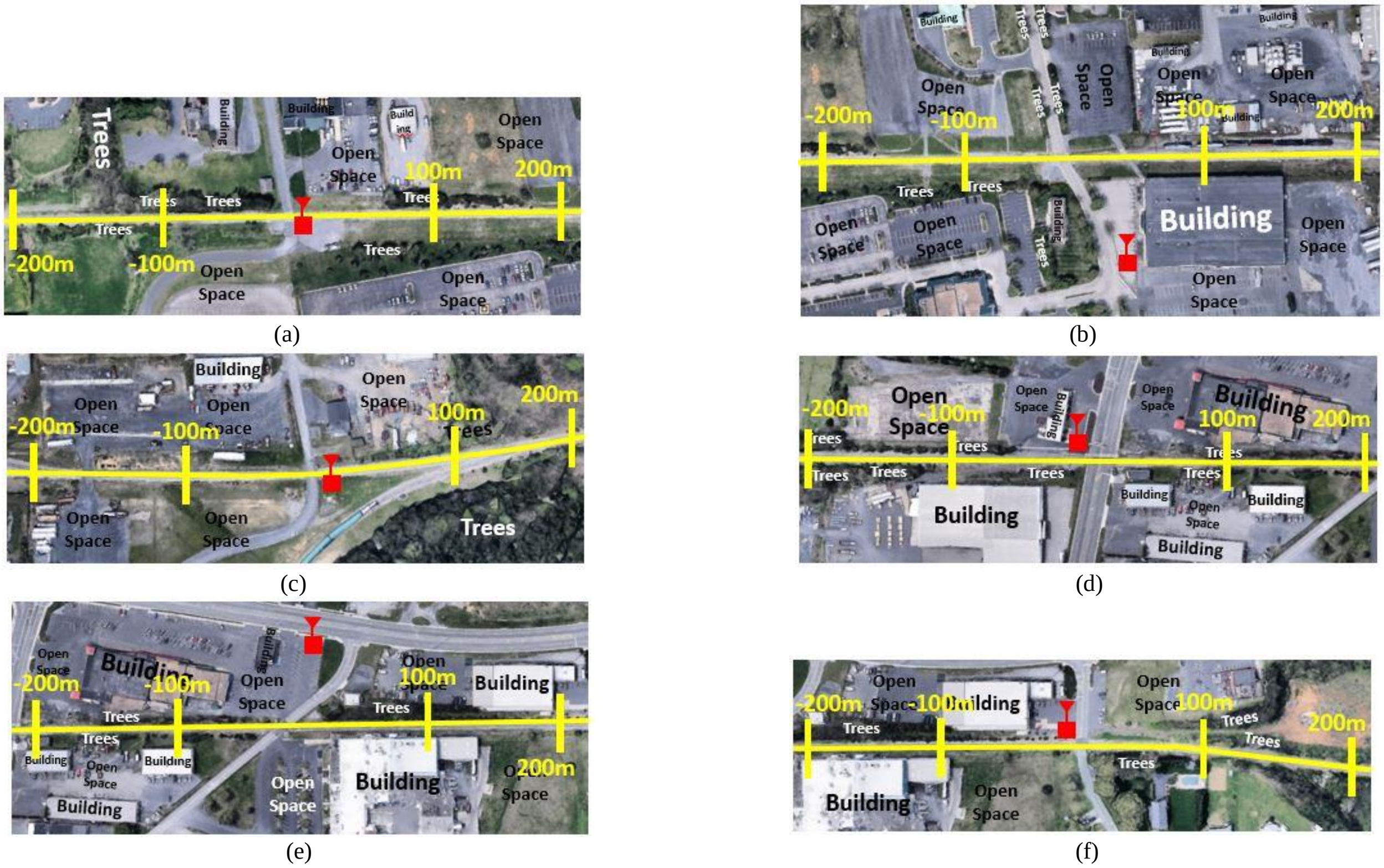

Fig. 11. DSRC PER results with respect to relative train position with center at the crossing with omnidirectional antenna and *indirect warning* scenario: (a) crossing #1, #3, and #6; (b) crossing #4

first crossing to 200 m after the last crossing; This is shown in Fig. 11.

## A. DSRC Performance Results

The effective antenna gains, omnidirectional and bidirectional, for the *indirect* and *direct warning* cases vary as a function of the train position due to the not perfectly matching transmit and receive with the main lobe, especially for a bidirectional antenna. With the assumption of no obstructions between the transmitter and the receiver, we calculate the antenna gains as a function of the train position for the antenna gain patterns shown in Fig. 8. The antenna gains for the omnidirectional and the bidirectional antennas for the *indirect* and *direct warning* are shown in Fig. 12. For performance assessment of the omnidirectional and bidirectional antennas,

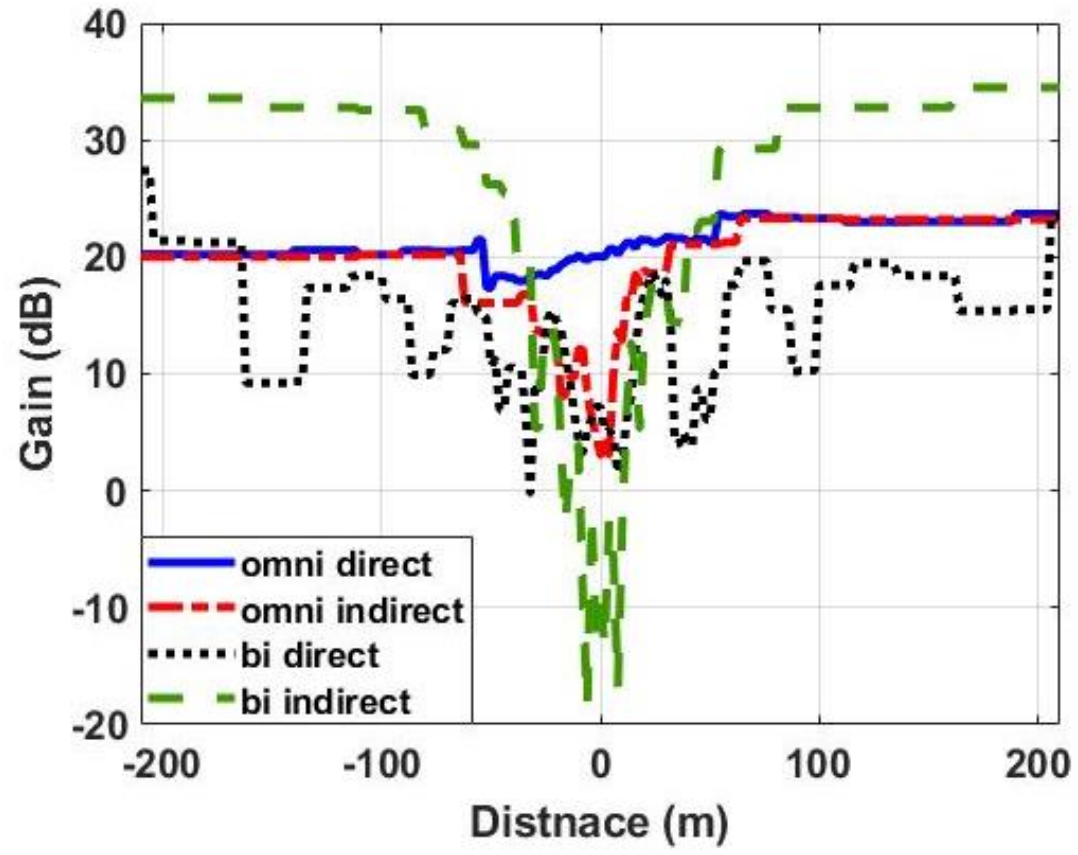

Fig. 12. Antenna gain for (a) omnidirectional direct scenario; (b) omnidirectional indirect scenario; (c) bidirectional direct scenario; (d) bidirectional indirect scenario

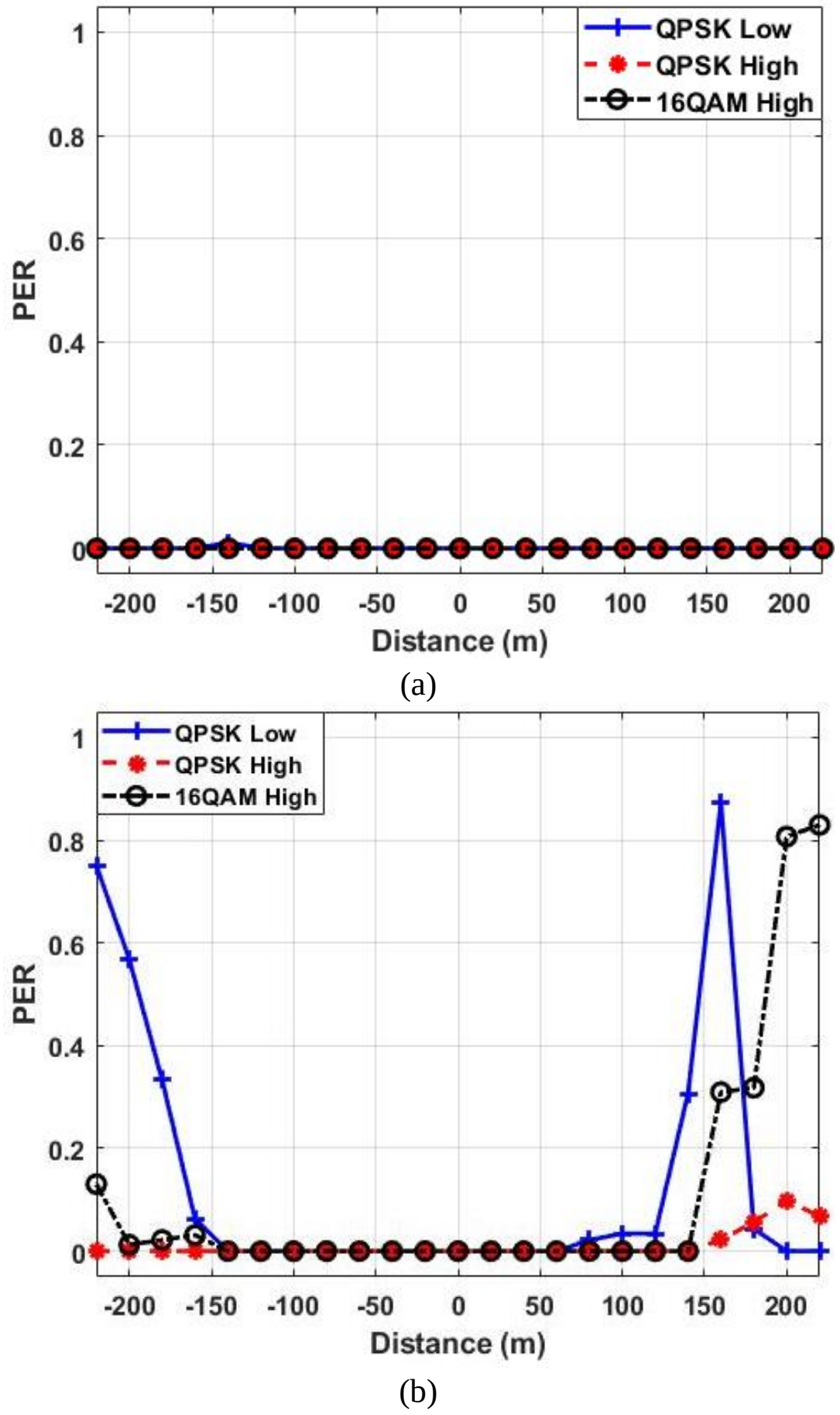

Fig. 13. DSRC PER results with respect to relative train position with center at the crossing with omnidirectional antenna and *indirect warning* scenario: (a) crossing #1, #3, and #6; (b) crossing #4

we consider the omnidirectional with high transmit power and bidirectional with low power in the *indirect warning* case. For the *direct warning* case, the performance of the bidirectional antenna is expected to be inconsistent due to the large effective antenna gain variations as seen in Fig. 12.

Fig. 13 shows the DSRC performance figures for the *indirect warning* case at crossings #1, #3, and #6 using an omnidirectional antenna on the train and the RSU. The PER is 0 in the range of 200 m before and after each crossing. These crossings assimilate rural propagation environments. Fig. 13(b) shows the results for crossing #4, where the PER is about 0 in the range of 150 m before and after the crossing for all radio configurations. The high power QPSK signaling provides excellent results even beyond this range. The surrounding of crossing #4 can be classified as suburban.

Fig. 14 shows the corresponding results when using a bidirectional antenna. From Fig. 14(a) we observe that the warning signal coverage ranges from 200 m before to 200 m after the crossing for crossings #1, #3, and #6, which are similar to the observation shown in Fig. 13(a). Also similar to the observation for the omnidirectional antenna experiment, the coverage range reduces with higher modulation type and lower transmission power for crossing #4 when using a bidirectional antenna (Fig. 14(b)).

For a fair comparison between the omnidirectional and bidirectional antenna systems, the coverage ranges of the omnidirectional antenna with QPSK and high transmission power and the bidirectional antenna with QPSK and low transmission power are compared. In rural settings, i.e. crossing #1, #3, and #6, for the range of interest, both antennas show the same performance. In a suburban-like setting, i.e. crossing #4, we observe that the omnidirectional antenna extends the coverage by 50 m after the crossing when compared to the bidirectional antenna.

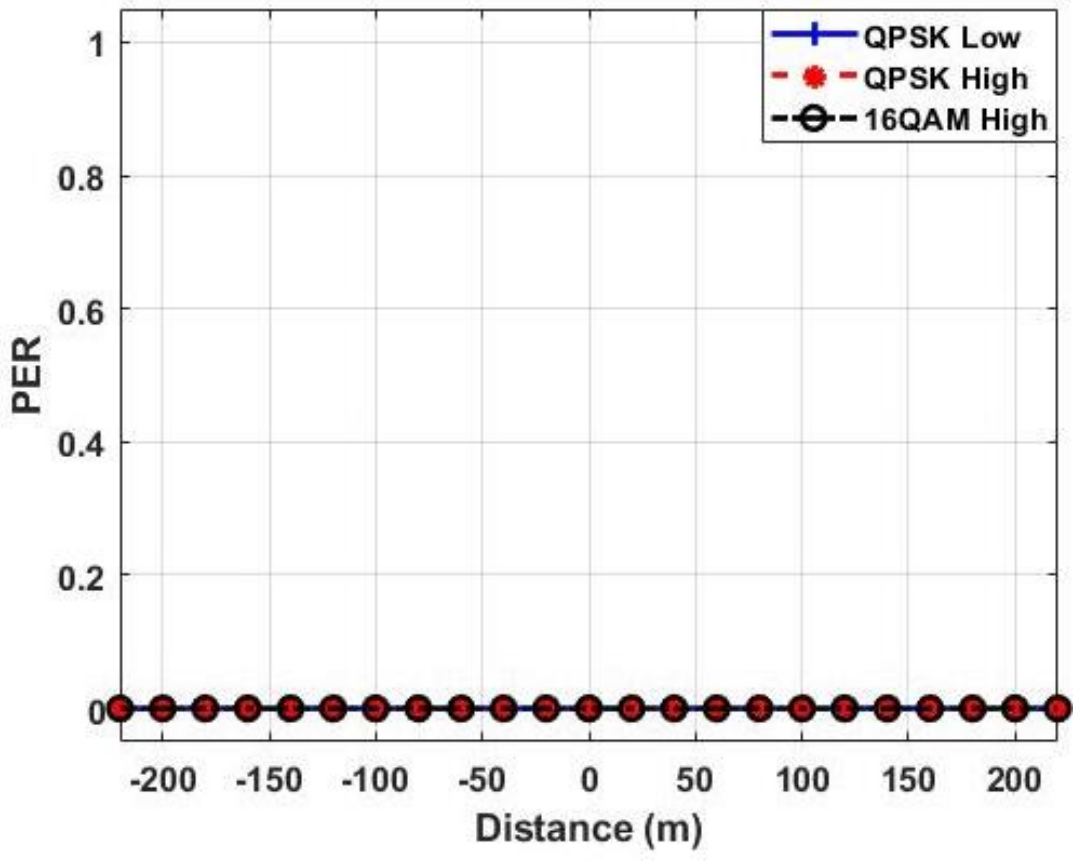

(a)

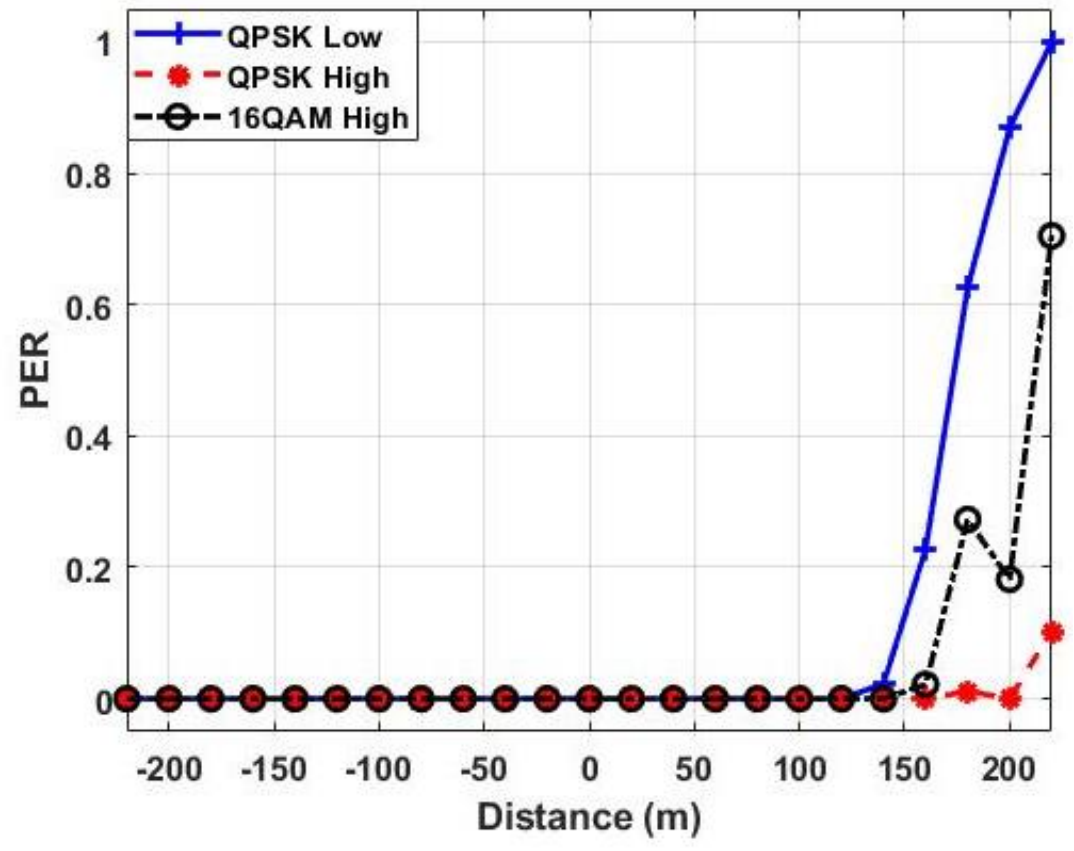

(b)

Fig. 14. DSRC PER results with respect to relative train position with center at the crossing with bidirectional antenna and *indirect warning* scenario: (a) crossing #1, #3, and #6; (b) crossing #4

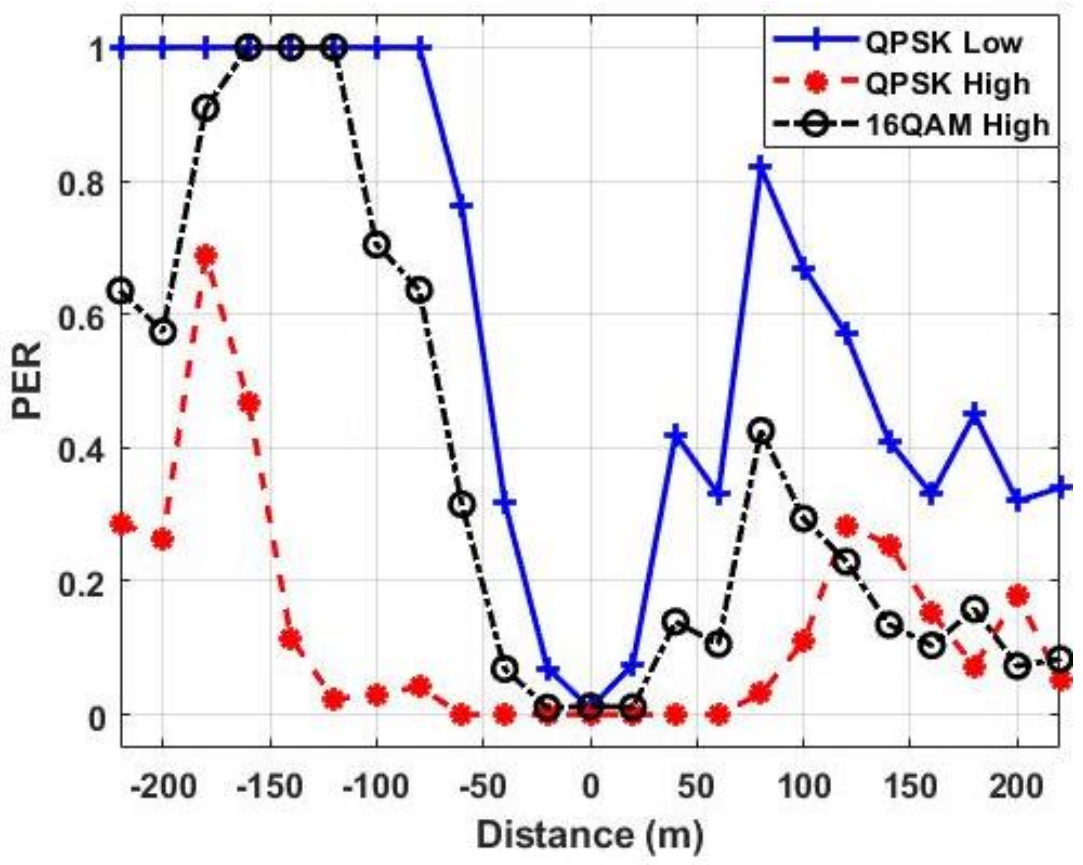

(a) **Same label as Fig. 15(a)**

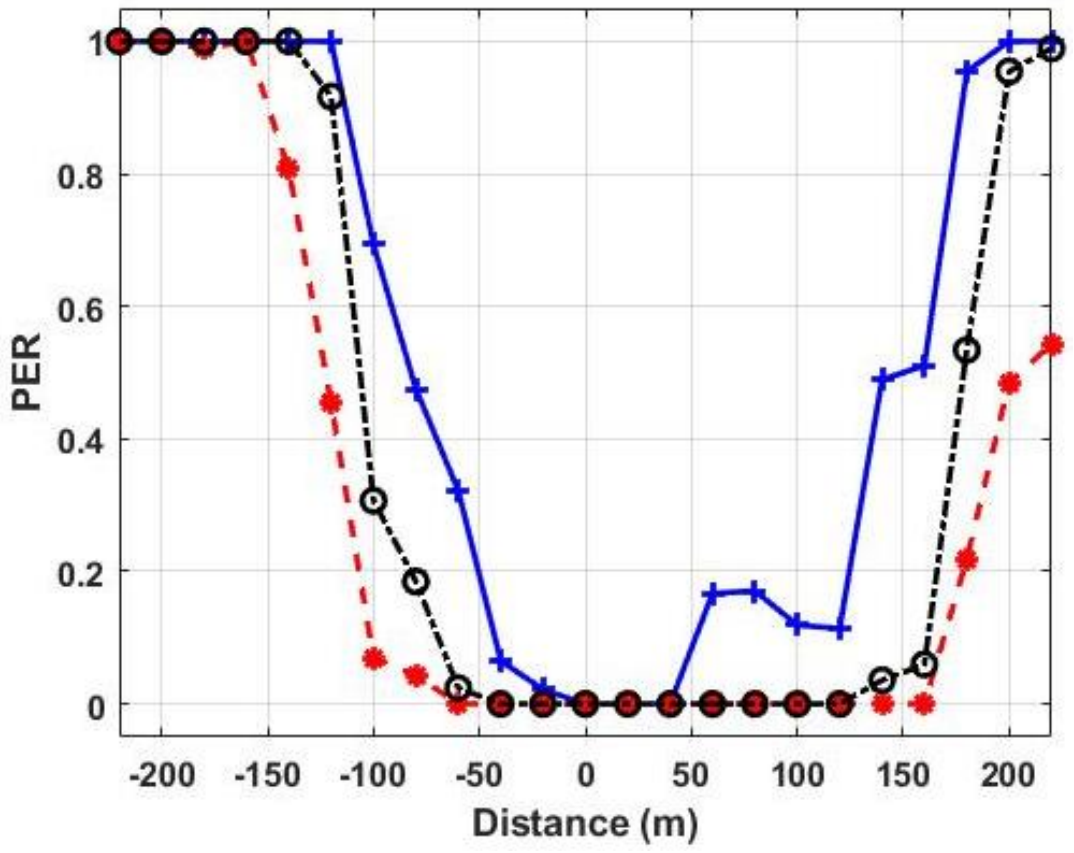

(b)

Fig. 15. DSRC PER results with respect to relative train position with center at the crossing with omnidirectional antenna and *direct warning* scenario: (a) crossing #2; (b) crossing #5

Fig. 15 presents the DSRC performance results for the *direct warning* case with an omnidirectional antenna. We observe that the window of good PER performance is much narrower than for the *indirect warning* cases. However, QPSK with high power can still provide more than 100 m warning coverage range before and after the crossing. Interestingly, the coverage range differences between 16QAM with high power and QPSK with low power is similar to what we observed for the *indirect warning* case; however, the differences between QPSK with high power and the other two configurations are less pronounced.

Fig. 16 presents the DSRC performance results for the *direct warning* case with a bidirectional antenna. Because of the inconsistent antenna gains (Fig. 12), the performances for crossings #2 and #5 are fluctuating. Despite the inconsistent performance, more than +/-100 m warning coverage range can be provided by the QPSK system with a high transmit power.

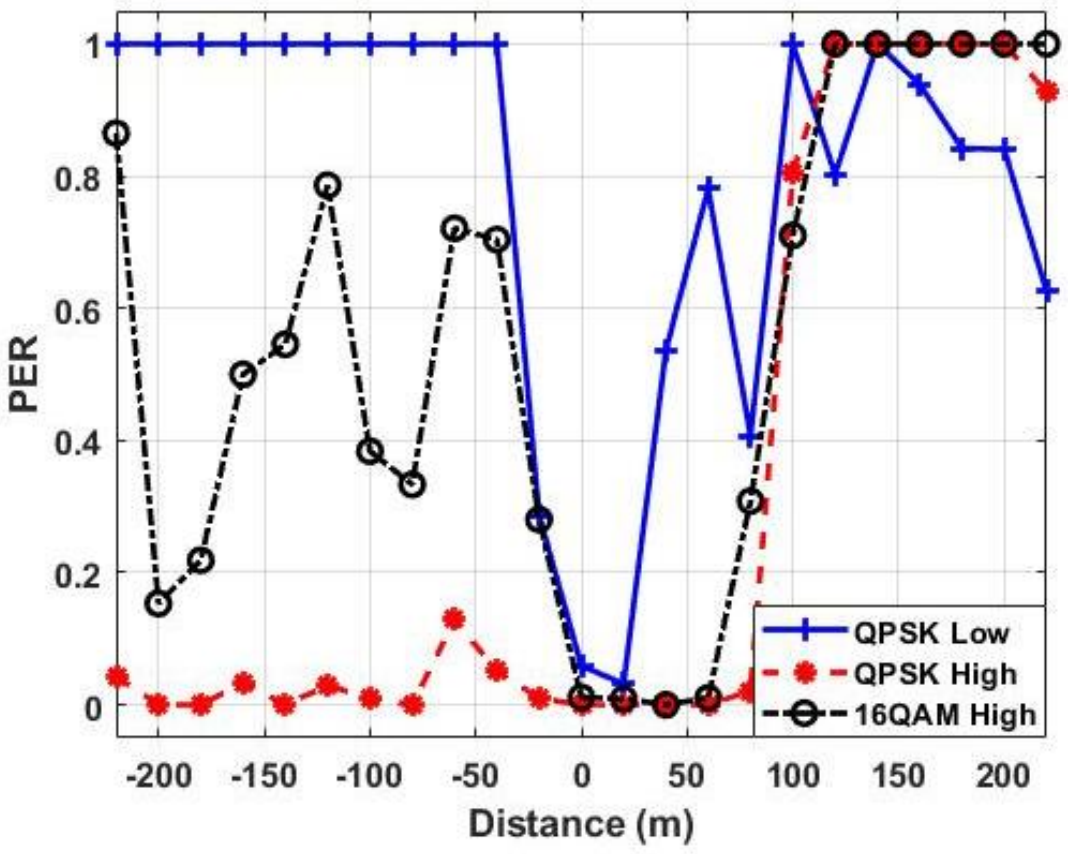

(a) **Same label as Fig. 16(a)**

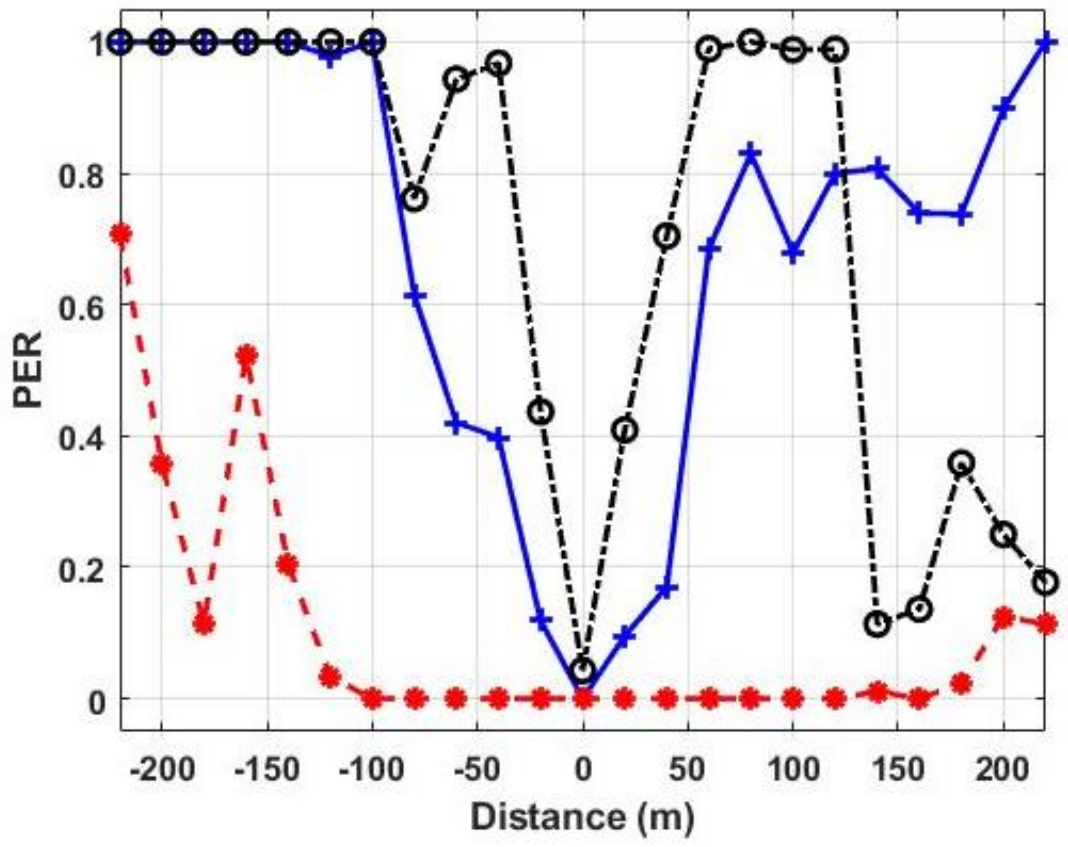

(b)

Fig. 16. DSRC PER results with respect to relative train position with center at the crossing with bidirectional antenna and *direct warning* scenario: (a) crossing #2; (b) crossing #5

Similar to the observations for the *indirect warning* case, the coverage range with bidirectional antenna, QPSK and low power transmission is narrower than with omnidirectional antenna, QPSK, high power transmission.

From these measurements, the omnidirectional and bidirectional antennas provide almost identical performances in a rural environment, especially where there are no buildings nearby. For the *indirect warning* case, both antenna systems are appropriate for the safety application. However, for the *direct warning* case, because of the inconsistent effective gains of the bidirectional antenna over the train's trajectory, the omnidirectional antenna provides more consistent performance. Even though the performances with different configurations are varying, since the warning application can operate at a higher PER than other commercial applications, the configuration that guarantees a wider coverage range is an omnidirectional antenna with QPSK modulation scheme and high power transmission. Moreover, the environment near the crossings is important to consider to ensure good system performance for the specific application.

## B. Latency Analysis

Latency is measured as the time from the packet generation at the source to the packet decoding at the sink; we use only successfully received packets. This latency comprises the packet processing at the transmitter, the radio propagation, and the packet processing at the receiver. It is $t_s$ in (3) and helps to analyze the early warning system by applying it into (4).

Fig. 17 represents the empirically measured latency for the SVRR tracks for the QPSK modulation scheme and high power scenario. The latencies are independent of the distances, as expected, because the propagation delay, $t_{prop}$, is negligible. We observe that it depends on the surrounding. For crossings #1, #2, #3, and #6, the latency values are more steady than for crossings #4 and #5, which are more suburban than the other crossings. Most of the packets are received within 5 ms. A few packets are delayed between 5 and 7.5 ms at crossings #4 and #5. These peaks appear in geometries where trees or buildings are near the track. The conducted measurements use only one transmitter and one receiver, the minimum network density. Most measured latency values are less than the transmission period of 50 ms. The transmission rate of BSM that our radio configuration setup is 20 Hz, that is, they a packet is transmitted every 50 ms.

Since the warning application is broadcasting signals from a single transmitter to multiple receivers, the latency of multiple identical receivers would be similar. Also, the measured latency from transmitter to receiver is less than 5 ms for both the *direct* and *indirect warning* cases. Since the measured latency values are almost consistent across all six crossings, we may consider $t_s$ as a constant value of 5 ms to evaluate the safeness level, $\psi$.

## C. Safeness Level Analysis

The protection time ($t_{prot}$) that the warning system can provide can be obtained by calculating the time difference between where the safeness level, $\psi$, is 0 and 1. The safeness level, $\psi$, can be evaluated using (4) which captures the time that the vehicle driver needs to notice the warning and react to it ($t_r$), the system delay ($t_s$), the braking time ($t_b$), the remaining time for train to reach the crossing ($t_t$), and the time to avoid the collision ($t_{TAC}$). Parameters $t_r$ and $t_b$ are known and from prior studies [21] and [28], respectively. Parameters $t_{TAC}$ and $t_s$ were obtained in Section VI. Since we found that the QPSK modulation with a high transmission power provides the widest coverage range in our measurements, only this scenario is considered to evaluate the safeness level. Fig. 18 plots the safeness level for the *indirect* and *direct warning* for 25-65 mph vehicle speeds with dry and wet road conditions based on actual measured data. The effects of different train speeds on the safeness level are shown in Fig. 19.

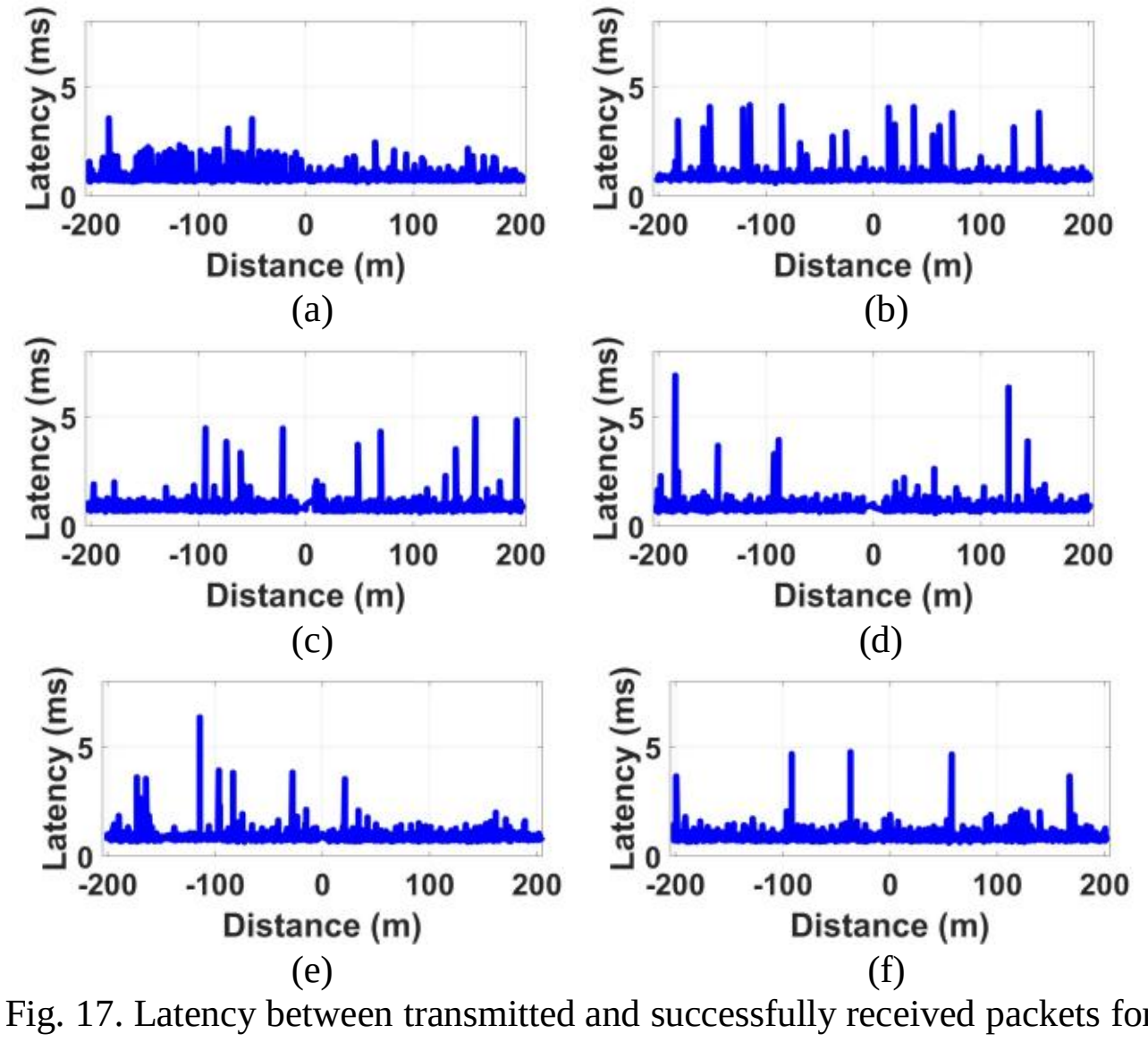

Fig. 17. Latency between transmitted and successfully received packets for omnidirectional with QPSK and high power scenario at (a) crossing #1, (b) crossing #2, (c) crossing #3, (d) crossing #4, (e) crossing #5, (f) crossing #6

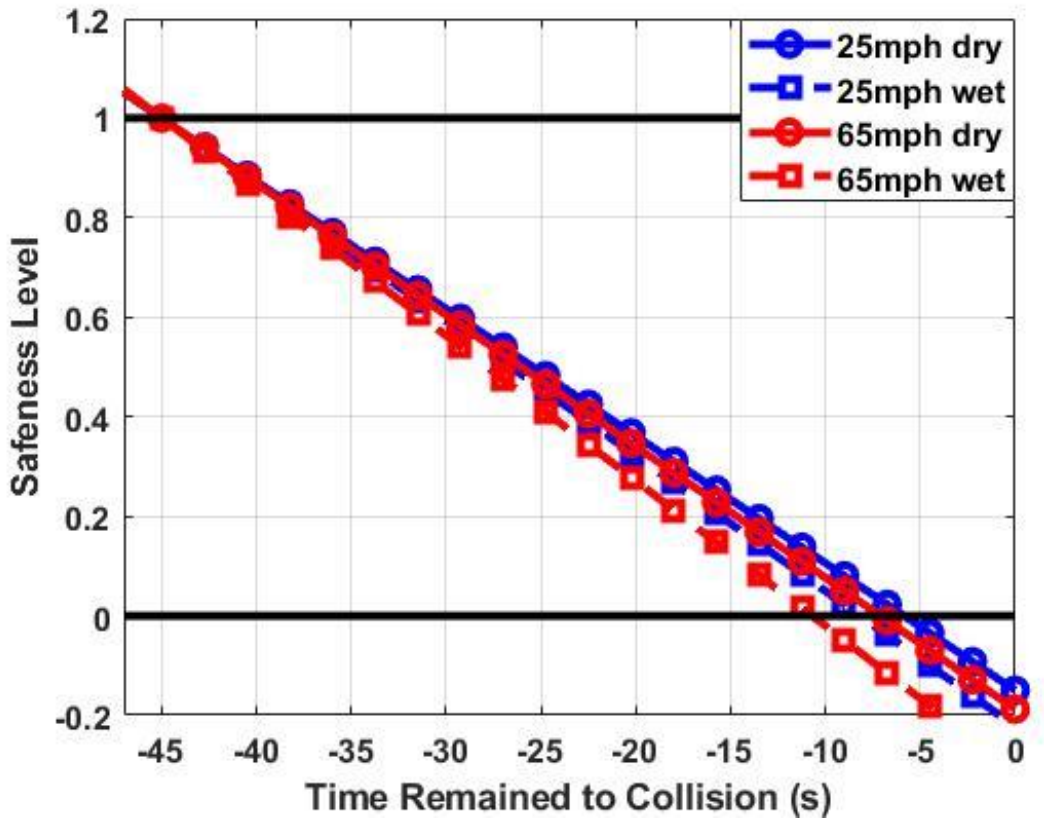

(a)

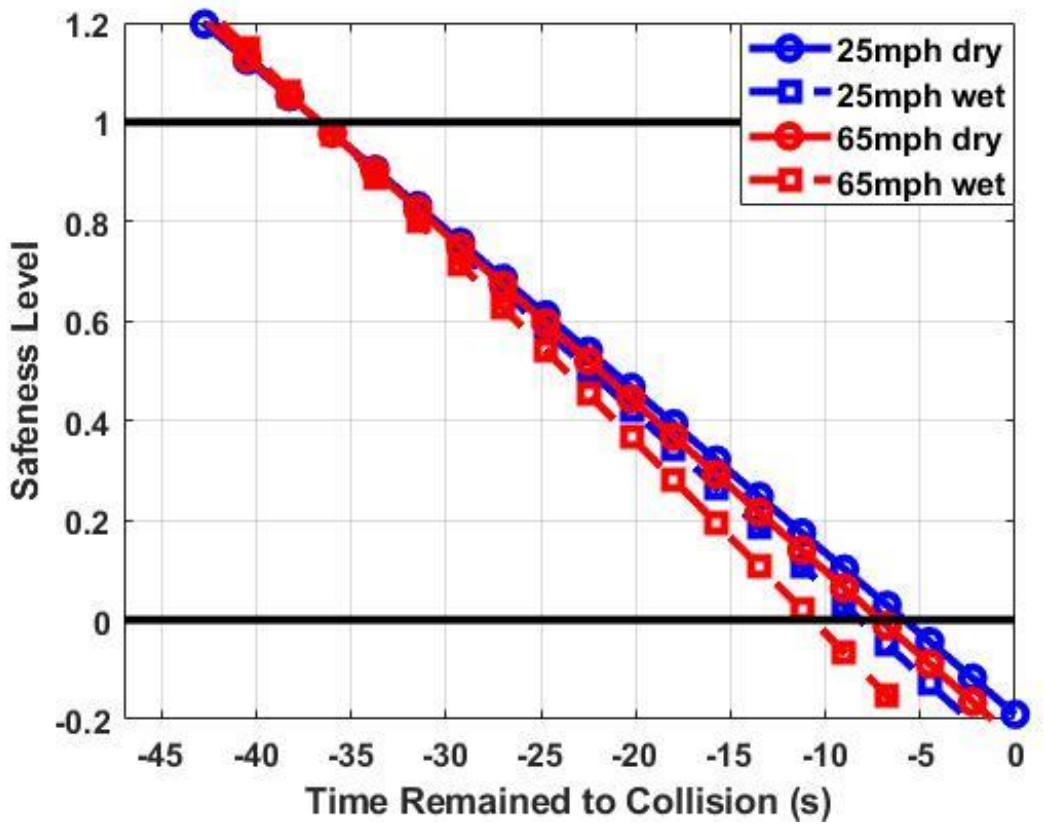

(b)

Fig. 18. Safeness level based on SVRR measurements for ; (a) *indirect warning*, (b) *direct warning*

As shown in Fig. 18(a), the system can provide 35-40 s of $t_{prot}$ for 25-65 mph vehicle speeds with dry and wet road conditions for the *indirect warning* case. The safeness level analysis uses the warning coverage range, $d_{warn}$, of 200 m, which is obtained from Fig. 13(a). However, the coverage range may be much larger than 200 m, possibly more than 500 m as shown in Fig. 9(a), and $t_{prot}$ may then be above 30 s. For the *direct warning* case, as shown in Fig. 18(b), the system can provide 25-30 s $t_{prot}$. Even though $t_{prot}$ differs by scenario, we can consider that a DSRC-enabled train safety communication system is feasible and useful for 10 mph train speed tracks because more than 25 s $t_{prot}$ can be provided by the warning system.

From the TTCI measurements, we found that the absolute number of correctly received packets is not dependent on train speed. The safeness level for different speeds is analyzed here with the help of Fig. 19. Therefore, we consider train speeds of 10, 25, and 35 mph with a $t_b$ of 2.3 s for 25 mph on dry road condition and 7.15 s for 65 mph on wet road condition with a fixed $d_{warn}$ of 200 m. The reduction of $t_{prot}$ is more affected by train speed than vehicle speed.

Since the evaluation is done with a $d_{warn}$ of 200 m, which is limited by the measurement setup, the available data cannot is insufficient to conclude about the system feasibility for trains that are faster than 35 mph. A feasibility analysis for faster trains needs to be evaluated for longer measurement ranges to obtain the definite coverage range. However, since $d_{warn}$ can provide more than 500 m in a wide-open space, as shown in Fig. 9(a), and the *indirect warning* case typically has a similar LoS channel as the wide-open space, the warning system is likely to be feasible for faster trains as well.

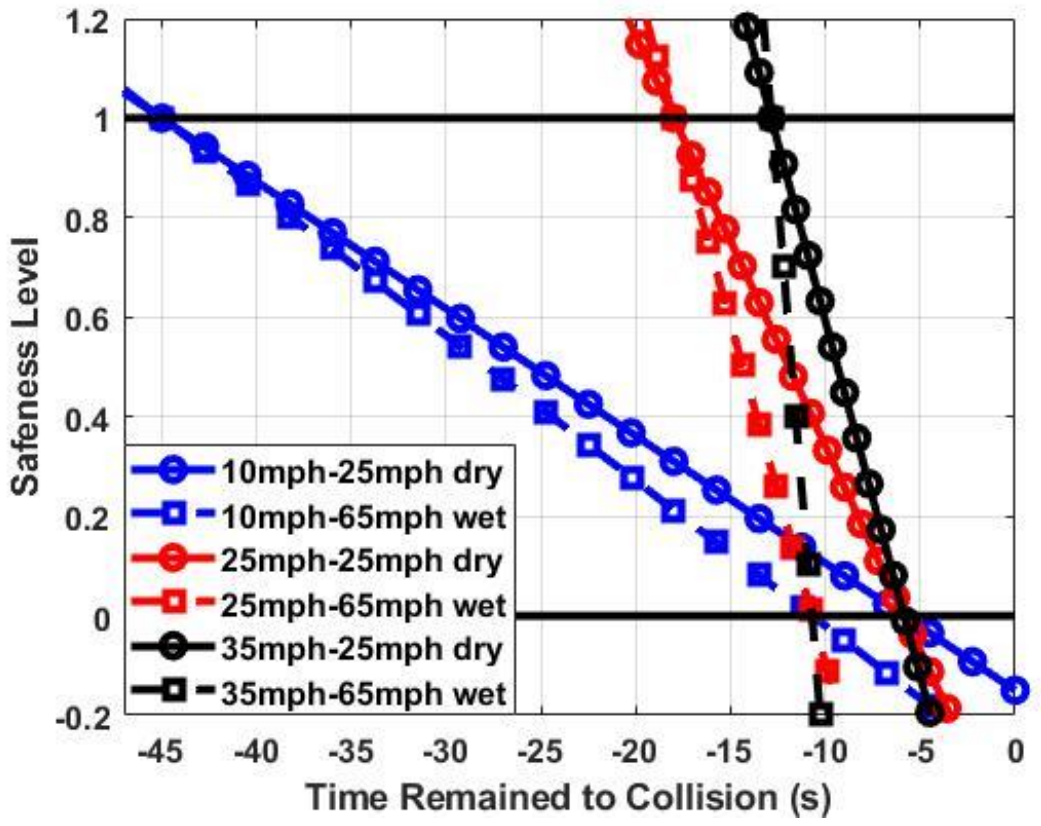

Fig. 19. Safeness level for different train speeds

## VII. Conclusions

Wireless communications protocols have been standardized and developed for safety-critical applications. These include DSRC and C-V2X. This study focuses on DSRC. Very few solutions have been presented for using DSRC for improving the safety near unprotected railroad crossings. The existence of train-to-vehicle accidents, suggests that effective warning systems to prevent train-to-vehicle collisions are needed. This paper presents the DSRC-enabled communications architecture for early warning with applications for unprotected crossings. Measurement data and analyses have been presented for two representative U.S. railroad tracks.

From the TTCI measurement evaluation, we conclude that the DSRC performance does not depend on the speed of the train. The numbers of correctly received packets are similar for the wide-open space and the artificial shadowing environments. The performance of the warning application depends on the operating environment.

From the SVRR measurements, we observe that the QPSK modulation scheme with an omnidirectional antenna and a high transmit power configuration can provide the widest coverage range. By processing the warning latency and distance-tagged radio performance results, we find that the system can provide 35-40 s protection time using the *indirect warning* case and 25-30 s for the *direct warning* case. The obtained protection time is large enough to consider the warning application feasible in a production environment in the U.S. Since the coverage range is independent of train speed, $t_{prot}$ is a function of the train speed rather than vehicle speed. Therefore, the minimum required range to evaluate a radio-based warning system before its implementation is the multiplication of the train speed and sum of reaction time, $t_r$, which is approximately 3.5 s, and the braking time, $t_b$, which is dependent on the vehicle speed as shown in Table I.

As a result of our measurements and analyses presented here, we conclude that a DSRC-enabled train safety communication system is feasible and useful, especially for unprotected railroad crossings. Moreover, we recommend placing one or more RSUs at strategic locations in case of heavily obstructed crossings. Such an *indirect warning* case increases reliability. We presented and evaluated a point-to-point communication system architecture, but the analyses can be directly extended to multiple radios and relays. We used commercial DSRC equipment for the measurements presented in this paper. Cellular-V2X (C-V2X), based on LTE and future 5G, is an alternative technology that has been introduced for V2X to improve road safety and traffic flow efficiency [32, 33]. Therefore, C-V2X is an alternative candidate for the proposed warning system, where the proposed safeness evaluation methodology can be applied to assess it for T2V and safety at railroad crossings.

## Acknowledgment

This work was supported by the Institute of Information & communications Technology Planning & Evaluation (IITP) grant funded by the Korea government (MSIT) (No. 2020-0-00839, Development of Advanced Power and Signal EMC Technologies for Hyper-connected E-Vehicle). This research was supported by the MSIT(Ministry of Science and ICT), Korea, under the ITRC(Information Technology Research Center) support program(IITP-2021-2016-0-00291) supervised by the IITP(Institute for Information & Communications Technology Planning & Evaluation)

## Referencesc

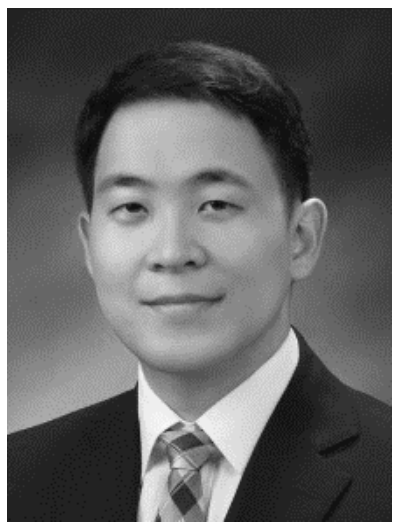

**Junsung Choi** (choijs89@kaist.ac.kr) is currently a researcher in the CCS Graduate school of Green Transportation with KAIST, Daejeon, Korea. He received B.S., M.S., and Ph.D. degree in Electrical and Computer Programming Engineering from Virginia Tech in 2013, 2016, and 2018, respectively. He served as a member of Wireless@VT between 2013 and 2018. During the M.S. and Ph.D. degree, he was a Research Assistant (the Bradley Department of Electrical and Computer Engineering, Virginia Tech). His research interests include vehicular communications, V2X communications, propagation channel characteristics, 5G communications, and LPI communications.

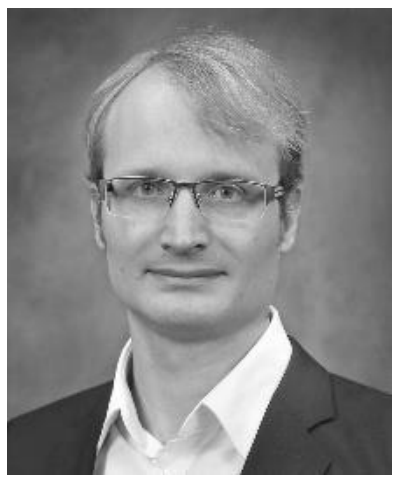

**Vuk Marojevic** (vuk.marojevic@ msstate.edu) is an associate professor in electrical and computer engineering at Mississippi State University. He obtained his M.S. from the University of Hannover, Germany, in 2003 and his PhD from Barcelona Tech-UPC, Spain, in 2009, both in electrical engineering. His research interests are in 4G/5G security, spectrum sharing, software radios, testbeds, resource management, and vehicular and aerial communications technologies and systems. He is an Editor of the IEEE Trans. on Vehicular Technology, an Associate Editor of the IEEE Vehicular Technology Magazine, and an Officer of the IEEE ComSoc Aerial Communications Emerging Technology Initiative.

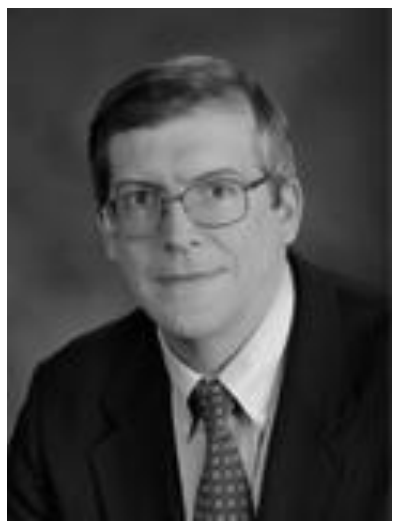

**Carl B. Dietrich** (cdietric@vt.edu) is a Research Associate Professor in the Bradley Department of Electrical and Computer Engineering at Virginia Tech. He earned Ph.D. and M.S. degrees in Electrical Engineering from Virginia Tech, Blacksburg, VA, and a B.S. in Electrical Engineering from Texas A&M University, College Station, TX. His research interests include cognitive radio, software defined radio, multi-antenna systems, and radio wave propagation. Dr. Dietrich has chaired the Wireless Innovation Forum's Educational Special Interest Group, is an IEEE Senior Member and a member of IEEE Eta Kappa Nu and ASEE. He is also a licensed professional engineer in Virginia.

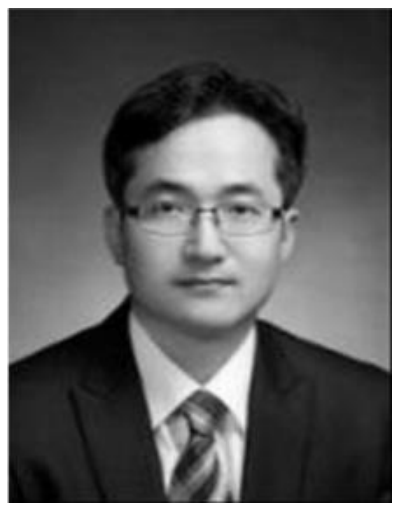

**Seungyoung Ahn** (sahn@kaist.ac.kr) received the B.S., M.S., and Ph.D. degrees from KAIST, Daejeon, South Korea, in 1998, 2000, and 2005, respectively. He is currently an Associate Professor with KAIST. His research interests include wireless power transfer (WPT) system design and electromagnetic compatibility design for electric vehicle and digital systems.